\documentclass[preprint,aps,prd,floatfix,nofootinbib,11pt]{revtex4-2}
\pdfoutput=1
\usepackage{graphicx}
\usepackage{amsfonts}
\usepackage{nicefrac}
\usepackage{xcolor}
\usepackage{natbib}
\usepackage{braket}

\usepackage{hyperref}

\usepackage{amsmath}
\usepackage{rotating}
\usepackage{amssymb,amsthm}
\usepackage{soul}

\usepackage{xcolor}
\usepackage[normalem]{ulem}

\usepackage{latexsym}
\usepackage{graphics}

\usepackage{amsfonts}
\usepackage{amsmath}
\usepackage{rotating}
\usepackage{amssymb}

\usepackage{amsmath}
\usepackage{rotating}
\usepackage{amssymb}
\usepackage{soul}

\usepackage{xcolor}

\usepackage{xcolor}
\def\beq{\begin{eqnarray}}  
	\def\eeq{\end{eqnarray}}

\def\beq{\begin{eqnarray}}  
\def\eeq{\end{eqnarray}}

\begin{document}

\title{Kantowski-Sachs spherically symmetric solutions in teleparallel $F(T)$ gravity}

\author{A. Landry}
\email{a.landry@dal.ca}
\affiliation{Department of Mathematics and Statistics, Dalhousie University, Halifax, Nova Scotia, Canada, B3H 3J5}

\begin{abstract}
In this paper, we investigate time-dependent Kantowski-Sachs spherically symmetric teleparallel $F(T)$ gravity in vacuum and in a perfect isotropic fluid. We begin by finding the field equations and solve for new teleparallel $F(T)$ solutions. With a power-law ansatz for the coframe functions, we find new non-trivial teleparallel $F(T)$ vacuum solutions. We then proceed to find new non-trivial teleparallel $F(T)$ solutions in a perfect isotropic fluid with both linear and non-linear equation of state. We find a great number of new exact and approximated teleparallel $F(T)$ solutions. These classes of new solutions are relevant for future cosmological applications.

\end{abstract}

\maketitle

\newpage

\tableofcontents

\newpage

\section{Introduction}

The teleparallel $F(T)$-type theory of gravity is a promising alternative theory to General Relativity (GR) \cite{Aldrovandi_Pereira2013,Bahamonde:2021gfp,Krssak:2018ywd}. This is a theory where the geometry is described by the spacetime torsion, a function of the coframe ${\bf h}^a$ (and its derivatives) and the spin-connection $\omega^a_{~bc}$. The teleparallel gravity mainly works with a frame basis instead of a metric tensor. The role of symmetry is not as clear as in pseudo-Riemannian geometry, where symmetry is defined in terms of the Killing Vectors (KVs). The Riemannian geometry in GR is defined by the Levi-Civita curvature and full dependent on the metric. But it is very different for the case of teleparallel $F(T)$-type gravity.

The frame based approach development for determining the spacetime symmetries has been explored \cite{chinea1988symmetries, estabrook1996moving, papadopoulos2012locally}. A possible complication arises from the possibility of non-trivial linear isotropy group: a Lie group of Lorentz frame transformations keeping the associated tensors of the geometry invariant. A new approach was introduced for symmetry determination of any geometry based on an independent frame and spin-connection admitting the torsion and curvature tensors as geometric objects \cite{MCH}. In this case, the spin-connection is an independent object and any geometry with a zero curvature and a zero non-metricity tensors is defined as a {\it teleparallel geometry}. The approach is based on the existence of a particular class of invariantly defined symmetry frames. For a non-trivial linear isotropy group, the symmetry determination needs to define a set of inhomogeneous differential equations (DEs) for coframes and spin-connection in an orthonormal gauge $g_{ab} = diag[-1,1,1,1]$. In addition, the spin-connection ${\omega}^a_{~bc}$ will be defined in terms of an arbitrary Lorentz transformation $\Lambda^a_{~b}$ and obtained from the zero curvature requirement (see ref.\cite{MCH} and references within). All these considerations can be summarized as \cite{olver1995equivalence}:
\begin{subequations}
\begin{align}
	\mathcal{L}_{{\bf X}} {\bf h}^a =& \lambda^a_{~b} {\bf h}^b , \label{intro1}
	\\
	 \mathcal{L}_{{\bf X}} {\omega}^a_{~bc} =& 0, \label{Intro2}
	\\
	\omega^a_{~bc} = & \Lambda^a_{~d} {\bf h}_c((\Lambda^{-1})^d_{~b}), \label{TP:con}
\end{align}
\end{subequations}
where ${\bf h}^a$ is the orthonormal coframe basis, $\mathcal{L}_{{\bf X}}$ is the Lie derivative with respect to the KV ${\bf X}$, $\lambda^a_{~b}$ is an invariant Lie algebra generator of Lorentz transformations $\Lambda^a_{~b}$.

A well-known subclass of teleparallel gravitational theories equivalent to GR is the Teleparallel Equivalent to General Relativity (TEGR), based on a torsion scalar $T$ constructed from the torsion tensor \cite{Aldrovandi_Pereira2013}. The most common generalization of TEGR is the teleparallel $F(T)$-type gravity, where $F$ is a function of the torsion scalar $T$ \cite{Ferraro:2006jd, Ferraro:2008ey, Linder:2010py}. In the covariant approach to $F(T)$-type gravity, the teleparallel geometry is gauge invariant with a zero curvature and a zero non-metricity spin-connection. This last quantity is zero in all proper frames and non-zero in all other frames \cite{Lucas_Obukhov_Pereira2009,Aldrovandi_Pereira2013,Krssak:2018ywd}. Therefore, the resulting teleparallel gravity theory is locally invariant with covariant FEs under Lorentz definition \cite{Krssak_Pereira2015}. A proper frame is not invariantly defined in terms of the connection (a non-tensorial quantity) and some problems arises when using such a frame to determine symmetries. However, some previous considerations have also been developed, used and/or adapted for theories close to teleparallel $F(T)$ theories. We can think of the New General Relativity (NGR) described essentially by the use of the three irreducible components of the torsion tensor (see refs. \cite{kayashi,beltranngr,bahamondengr} and references within). Some of the previous elements were also  considered for the symmetric teleparallel $F(Q)$-type gravity, a theory in development with some potential (see refs. \cite{heisenberg1,heisenberg2,faithman1,hohmannfq} and references within). Or again, some of these elements are also useful for improving the study of the geometrical trinity of gravity and some intermediate theories like $F(T,Q)$-type, $F(R,Q)$-type, $F(R,T)$-type and others (see refs. \cite{jimeneztrinity,nakayama,ftqgravity,ftqspecial,frqspecial,frtspecial,frttheory} and references within). These examples show a some number of possible approaches, but the best is to go first with the teleparallel $F(T)$-type approach.

There is an important number of papers in the literature on spherically symmetric solutions in teleparallel $F(T)$ gravity \cite{golov1,golov2,golov3,debenedictis,SSpaper,TdSpaper,baha1,bahagolov1,awad1,baha6,nashed5,pfeifer2,elhanafy1,benedictis3,baha10,baha4,calza}. The main features can be summarized by power-law $F(T)$ solutions obtained from power-law frame components (see \cite{golov1,golov2,golov3,debenedictis} and references within). Most of these papers use the Weitzenback gauge (proper frames satisfying trivially the antisymmetric FEs), but extra degrees of freedom (DoF) arise by imposing the zero spin-connection. This specificity leads to only symmetric parts of FEs and the presented $F(T)$ solutions are essentially limited to power-law for a big coframe expression. These symmetric FEs and the solutions are similar between the different gauges, but performing a frame changing is necessary for solving this extra DoF potential issue. When a non-proper frame is used, the non-zero spin connections are solutions to the non-trivial antisymmetric parts of FEs, and then all DoFs are covered by the FEs. This method is used in a paper on Teleparallel spherically symmetric geometries focusing on vacuum solutions and additional symmetry structures \cite{SSpaper}. The general FEs are defined in an orthonormal gauge assuming a diagonal frame and a non-trivial spin-connection, leading to non-trivial antisymmetric and symmetric parts of FEs without extra DoF. This paper also studied the Kantowski-Sachs (KS) geometry case and found the vacuum $F(T)$ solutions by using a special power-law ansatz. Therefore we need to find more solutions and going further for the KS spacetime, as done recently for static spacetime $F(T)$ solutions for perfect fluids \cite{nonvacSSpaper}. The KS spacetime fourth symmetry is defined by the radial-coordinate derivative $\partial_r$ leading to time-coordinate dependence for coframes, spin-connections and FEs.

In the literature, there are some works on KS spacetimes and solutions in GR and some specific $f(R)$-type theory of gravity  \cite{leon1,KScurvature,KSanisotropic}. In these papers, they carry out for some $F(R)$ solutions a detailed study concerning critical points, limits on  physical quantities, asymptotes and also the evolution of curvature to name a few. There are several other paper on more elaborated KS spacetime models, but they are not made in terms of teleparallel gravity and they are essentially focusing on $f(R)$-type gravity. For KS teleparallel $F(T)$ theory, there are a small number of recent works \cite{Rodrigues2015,Amir2015}. There are recently some generalizations of KS models and solutions for $F(T,B)$ gravity, the LRS Bianchi III Universe and even the $F(T,R)$-type and teleparallel $F(Q)$-type gravity. However, these papers focus essentially on cosmological solutions and possible equilibrium points of these cosmological systems \cite{leon2,leon3,frtkssol1,frtkssol2,fqks1,fqks2}. Therefore, their approaches do not really provide new class of teleparallel $F(T)$-type solutions. All these works relate that KS spacetimes geometries solutions are relevant and used for more refined cosmological solutions. Recently, there are some works on teleparallel KS solutions from Paliathanasis which may also lead to scalar field and quantized cosmological solutions \cite{palia2022KS,palia2023KS}. However no quantized solutions will be considered in the current paper, because the aim of the current paper is only to find $F(T)$ solutions for KS spacetimes without scalar field and quantization.

For this paper, we assume a time coordinate dependent spherically symmetric teleparallel geometry (a Kantowski-Sachs teleparallel geometry) in an orthonormal gauge as defined in ref \cite{SSpaper}. We will first find vacuum $F(T)$ solutions, then we will focus on finding perfect fluids Kantowski-Sachs teleparallel $F(T)$ solutions. After a summary of the teleparallel FEs and Kantowski-Sachs class of geometries in section \ref{sect2}, we will find in section \ref{sect3} additional  $F(T)$ solutions in vacuum. We then repeat the exercise in sect \ref{sect4} with linear equation of state (EoS). In section \ref{sect5}, we will solve FEs and find several approximated $F(T)$ solutions for a perfect fluid with a non-linear EoS. This paper has some common features, aims and structure with the paper on static perfect fluids teleparallel $F(T)$ solutions studied in ref. \cite{nonvacSSpaper}.

The notation is defined as: coordinate indices $\mu, \nu, \ldots$, tangent space indices $a,b,\ldots$ (see ref \cite{MCH}), spacetime coordinates $x^\mu$, frame fields ${\bf h}_a$ and its dual one-forms ${\bf h}^a$, vierbein $h_a^{~\mu}$ or $h^a_{~\mu}$, spacetime metric $g_{\mu \nu}$, Minkowski tangent space metric $\eta_{ab}$, spin-connection one-form $\omega^a_{~b} = \omega^a_{~bc} {\bf h}^c$, curvature tensor $R^a_{~bcd}$, torsion tensor $T^a_{~bc}$. The derivatives with respect to (w.r.t.) $t$ $F_{t} = F'$ (with a prime).

\section{Teleparallel spherically symmetric spacetimes and field equations}\label{sect2}

\subsection{Summary of teleparallel field equations}

\noindent The teleparallel $F(T)$-type gravity action integral is \cite{Aldrovandi_Pereira2013,Bahamonde:2021gfp,Krssak:2018ywd,SSpaper,nonvacSSpaper}:
\begin{equation}\label{1000}
S_{F(T)} = \int\,d^4\,x\,\left[\frac{h}{2\kappa}\,F(T)+\mathcal{L}_{Matter}\right],
\end{equation}
where $h$ is the coframe determinant and $\kappa$ is the coupling constant. We apply the least-action principle to the eqn \eqref{1000}, the symmetric and antisymmetric parts of FEs are \cite{SSpaper,nonvacSSpaper}:
\begin{subequations}
\begin{eqnarray}
\kappa\,\Theta_{\left(ab\right)} &=& F_T\left(T\right) \overset{\ \circ}{G}_{ab}+F_{TT}\left(T\right)\,S_{\left(ab\right)}^{\;\;\;\mu}\,\partial_{\mu} T+\frac{g_{ab}}{2}\,\left[F\left(T\right)-T\,F_T\left(T\right)\right],  \label{1001a}
\\
0 &=& F_{TT}\left(T\right)\,S_{\left[ab\right]}^{\;\;\;\mu}\,\partial_{\mu} T, \label{1001b}
\end{eqnarray}
\end{subequations}
with $\overset{\ \circ}{G}_{ab}$ the Einstein tensor, $\Theta_{\left(ab\right)}$ the energy-momentum, $T$ the torsion scalar, $g_{ab}$ the gauge metric, $S_{ab}^{\;\;\;\mu}$ the superpotential (torsion dependent) and $\kappa$ the coupling constant. The canonical energy-momentum is obtained from $\mathcal{L}_{Matter}$ term of eqn \eqref{1000} as:
\begin{align}\label{1001ca}
\Theta_a^{\;\;\mu}=\frac{1}{h} \frac{\delta \mathcal{L}_{Matter}}{\delta h^a_{\;\;\mu}}.
\end{align}
The eqn \eqref{1001ca} antisymmetric and symmetric parts are \cite{SSpaper}:
\begin{equation}\label{1001c}
\Theta_{[ab]}=0,\qquad \Theta_{(ab)}= T_{ab},
\end{equation}
where $T_{ab}$ is the symmetric part of energy-momentum tensor. The eqn \eqref{1001c} is valid only when the matter field interacts with the metric $g_{\mu\nu}$ defined from the coframe $h^a_{\;\;\mu}$ and the gauge $g_{ab}$, and is not directly coupled to the $F(T)$ gravity. This consideration is valid in the situation of zero hypermomentum (i.e. $\mathfrak{T}^{\mu\nu}=0$) as discussed in refs. \cite{golov3,nonvacSSpaper}. The hypermomentum is defined from eqns \eqref{1001a} and \eqref{1001b} components as \cite{golov3}:
\begin{align}\label{1001h}
\mathfrak{T}_{ab}=\kappa\Theta_{ab}-F_T\left(T\right) \overset{\ \circ}{G}_{ab}-F_{TT}\left(T\right)\,S_{ab}^{\;\;\;\mu}\,\partial_{\mu} T-\frac{g_{ab}}{2}\,\left[F\left(T\right)-T\,F_T\left(T\right)\right].
\end{align}
The conservation of energy-momentum in teleparallel gravity for $\mathfrak{T}^{\mu\nu}=0$ case states that $\Theta_a^{\;\;\mu}$ satisfies the relation \cite{Aldrovandi_Pereira2013,Bahamonde:2021gfp}:
\begin{align}\label{1001e}
\overset{\ \circ}{\nabla}_{\nu}\left(\Theta^{\mu\nu}\right)=0,
\end{align}
with $\overset{\ \circ}{\nabla}_{\nu}$ the covariant derivative and $\Theta^{\mu\nu}$ the conserved energy-momentum tensor. This eqn \eqref{1001e} is also the GR conservation of energy-momentum expression. Satisfaction of eqn \eqref{1001e} is automatically required because the $\mathfrak{T}^{\mu\nu}=0$ condition (null hypermomentum). For non-zero hypermomentum situations (i.e. $\mathfrak{T}^{\mu\nu}\neq 0$), we need to satisfy more complex conservation equations than eqn \eqref{1001e} (see ref. \cite{golov3} for details).

\noindent For a perfect and isotropic fluid with any EoS, the matter tensor $T_{ab}$ is \cite{hawkingellis1,coleybook,nonvacSSpaper}:
\begin{align}\label{1001d}
T_{ab}= \left(P(\rho)+\rho\right)\,u_a\,u_b+g_{ab}\,P(\rho),
\end{align}
where $P(\rho)=P(\rho(t))$ is the EoS in terms of the time-dependent fluid density $\rho=\rho(t)$ and $u_a=(-1,\,0,\,0,\,0)$.

\subsection{Spherically symmetric teleparallel Kantowski-Sachs geometry}

The orthonormal time-dependent Kantowski-Sachs resulting vierbein is \cite{SSpaper}:
\beq h^a_{~\mu} = Diag\left[  1,\, A_2(t),\,A_3(t),\,A_3(t) \sin(\theta) \right], \label{VB:SS} \eeq

\noindent where we are able to choose new coordinate such that $A_1(t)=1$ without any lost of generality. This will allow us to find cosmological-like solutions.

The spin-connection $\omega_{abc}$ components for time-dependent spacetimes is \cite{SSpaper}:
\beq \begin{aligned} & \omega_{341} = W_1(t),\quad\quad\quad\quad\quad &  \omega_{342} = W_2(t), \quad\quad\quad\quad\quad & \omega_{233} = \omega_{244} = W_3(t),
\\
& \omega_{234} = -\omega_{243} = W_4(t), \quad &  \omega_{121} = W_5(t),\quad\quad\quad\quad\quad & \omega_{122} = W_6(t), 
\\
& \omega_{133} = \omega_{144} = W_7(t),\quad\quad & \omega_{134} = -\omega_{143} = W_8(t), \quad & \omega_{344} = - \frac{\cos(\theta)}{A_3 \sin(\theta)}. \end{aligned} \label{Con:SS} \eeq

For eqns. \eqref{VB:SS} and \eqref{Con:SS}, the curvature vanishing requirement implies that the functions $W_i(t)$ must take the form: 
\beq \begin{aligned} 
W_1 &= -\chi',\quad  & W_2 = 0, \quad & W_3 = \frac{\cosh(\psi)\cos(\chi)}{A_3}, \quad & W_4 = \frac{\cosh(\psi)\sin(\chi)}{A_3},\\
W_5 &= -\psi', \quad & W_6 = 0, \quad & W_7 = \frac{\sinh(\psi) \cos(\chi)}{A_3},\quad & W_8 = \frac{\sinh(\psi) \sin(\chi)}{A_3},  
\end{aligned} \label{SS:TPcon} \eeq

\noindent where $\chi$ and $\psi$ are arbitrary functions of the coordinate $t$ ($\chi'=\chi_t$ and $\psi'=\psi_t$).

\subsection{Teleparallel Kantowski-Sachs Field Equations}

The antisymmetric part of the $F(T)$ FEs are \cite{SSpaper}:
\begin{align}\label{1020}
 \frac{F_{TT}(T)\, T' \cosh(\psi) \cos(\chi)}{A_3} &= 0, \quad\quad  \frac{F_{TT}(T)\,T' \sinh(\psi) \sin(\chi)}{A_3} =0. 
\end{align}
These eqns \eqref{1020} lead to $\psi = 0$ and $\chi = \frac{\pi}{2}$ (and also $\frac{3\pi}{2}$) for $T \neq$ constant as solution. By substituting eqn \eqref{1020} solution into eqn \eqref{SS:TPcon}, we find that $W_4=\delta=\pm 1$ ($W_j=0$ for $j\neq 4$) and then eqn \eqref{Con:SS} becomes:
\begin{align}
\omega_{234} = -\omega_{243} = \delta, \quad \omega_{344} = - \frac{\cos(\theta)}{A_3 \sin(\theta)}.
\end{align}

The torsion scalar and the symmetric FEs components are exactly for $\chi = \frac{\pi}{2}$ ($\delta=+1$) \cite{SSpaper}:
\begin{subequations}
\begin{align}
 T =& 2\left(\ln(A_3)\right)'\left(\left(\ln(A_3)\right)'+2\left(\ln(A_2)\right)'\right) - \frac{2}{A_3^2} , \label{301a}
\\
 B' =&  -\left(\ln\left(A_2\,A_3^2\right)\right)'+\frac{\frac{1}{A_3^2} - \left(\ln\left(\frac{A_2}{A_3}\right)\right)''}{\left(\ln\left(\frac{A_2}{A_3}\right)\right)' }, \label{301d} 
\\
 \kappa \rho + \frac{F(T)}{2} =& \left( T+\frac{2}{A_3^2} \right) F_T(T), \label{301b}
 \\ 
 -\kappa ( \rho + P) =& \left[\left(\ln(A_3)\right)'\left(B'+ \left(\ln(A_3)\right)'-\left(\ln(A_2)\right)'\right)+ \left(\ln(A_3)\right)''\right] F_T(T), \label{301c}
\end{align}
\end{subequations}
\noindent where $F_T(T) \neq$ constant and $B'=\partial_t\left(\ln\,F_T(T)\right)$. Comparing with the version of the Kantowski-Sachs $F(T)$-gravity FEs in the literature \cite{Rodrigues2015,Amir2015}, the FEs are different. For $\delta=-1$ FEs set, there are some small minor differences for some terms in eqns \eqref{301d} to \eqref{301c}, mainly some different signs at very specific terms. For the rest, the general form of the eqns \eqref{301a} to \eqref{301c} remains identical, regardless of $\delta$.

Then the conservation law for non-null $\rho$ and $P$ for time-dependent spacetimes is \cite{SSpaper}:
\begin{align}
\left(P+\rho\right)\,\left(\ln(A_2\,A_3^2)\right)'+\rho'=0, \label{302d}
\end{align}
where $\rho'=\rho_t$. For coming steps, we will solve eqns \eqref{301a} to \eqref{302d} and the solutions will also depend on the EoS, the $P(\rho)$ relationship.


\section{Vacuum solutions}\label{sect3}

By setting $P=\rho=0$ in eqns \eqref{301a} to \eqref{301c}, we obtain the symmetric FEs:
\begin{subequations}
\begin{align}
 T =& 2\left(\ln(A_3)\right)'\left(\left(\ln(A_3)\right)'+2\left(\ln(A_2)\right)'\right) - \frac{2}{A_3^2} , \label{351d}
\\
 B' =&  -\left(\ln\left(A_2\,A_3^2\right)\right)'+\frac{\frac{1}{A_3^2} - \left(\ln\left(\frac{A_2}{A_3}\right)\right)''}{\left(\ln\left(\frac{A_2}{A_3}\right)\right)' }, \label{351a}
\\
 F(T) =& 2\left( T + \frac{2}{A_3^2} \right) F_T(T), \label{351b}
\\
 0 =& \left[\left(\ln(A_3)\right)'\left(B'+ \left(\ln(A_3)\right)'-\left(\ln(A_2)\right)'\right)+ \left(\ln(A_3)\right)''\right]. \label{351c}
\end{align}
\end{subequations}
In this case, conservation laws are trivially satisfied because null fluid density and pressure. In ref \cite{SSpaper}, we solved the FEs described by eqns \eqref{351d} to \eqref{351c} by using the special ansatz $A_2=A_3^n$ where $n$ is a real number. We found a linear $A_3$ in time-coordinate $t$ and a pure power-law for $F(T)$ as solution with this specific ansatz. Some additional solutions are possible and we first use power-law ansatz for finding some of them for $A_2$, $A_3$ and $F(T)$. This type of coframe solutions is frequently used in recent literature and is easy to apply \cite{golov1,golov2,golov3,debenedictis,SSpaper,TdSpaper}. But above all, this approach has the advantage of easily distinguishing $F(T)$ teleparallel solutions from other types of solutions (such as GR solutions). In addition, this approach subsequently makes it possible to generalize with the exponential ansatz (an infinite sum of power-law terms: an appropriate limit case.) as well as to refine the coframe and $F(T)$ solutions. For these purposes, we will then focus on more specific possible solutions as exponential ansatz to name only this one.

\subsection{Power-law solutions}\label{sect31}

\noindent We will set the following power-law ansatz:
\begin{align}\label{350}
A_2=t^b\quad\quad\text{and}\quad\quad A_3=c_0\,t^c,
\end{align}
where $b_0=1$ because we can perform a coordinate transformation $d\tilde{r}=b_0\,dr$ for $A_2(t)$ frame component. The eqns \eqref{351d} to \eqref{351c} become:
\begin{subequations}
\begin{align}
T &= \frac{2c\left(c+2b\right)}{t^2} - \frac{2}{c_0^2\,t^{2c}} , \label{352d}
\\
t\,B' &= (1-b-2c) + \frac{t^{2-2c}}{c_0^2\,(b-c)}  , \label{352a}
\\
F(T) &= 2\left( T + \frac{2}{c_0^2\,t^{2c}} \right) F_T(T), \label{352b}
\\
0 &= c\left( t\,B' + (c-b-1)  \right) . \label{352c}
\end{align}
\end{subequations}
If $c=0$, eqn \eqref{352d} leads to constant torsion scalar: a GR solution. For $c\neq 0$, we can put together eqns \eqref{352a} and \eqref{352c} leading to the simplified relation:
\begin{align}
0 &=  -c_0^2\,(b-c)(2b+c) + t^{2(1-c)} .  \label{352e}
\end{align}
The only possible $t$ independent solution is $c=1$ leading to $A_3=c_0\,t$. Then eqn \eqref{352e} becomes:
\begin{align}
0 &= b^2-\frac{b}{2}-\left(\frac{1}{2}+\frac{1}{2c_0^2}\right) ,
\nonumber\\
&\Rightarrow\; b=\frac{1+k}{4},\label{352f}
\\
&\Rightarrow\; A_2=t^{\frac{1+k}{4}} , \label{352g}
\end{align}
where $k=\delta_1\sqrt{9+\frac{8}{c_0^2}}$ and $\delta_1=\pm 1$ ($k<-3$ and $k>3$ for real values). By substituting $c=1$ and eqn \eqref{352f}, we find from eqn \eqref{352d} the relation $t(T)$:
\begin{align}\label{353}
t^{-2}(T) = \frac{4\,T}{(3+k)(7-k)} .
\end{align}
By substituting eqn \eqref{353} into eqn \eqref{352b}, we find and solve the DE for $F(T)$:
\begin{align}\label{354}
\frac{F_T(T)}{F(T)}=&\left(\frac{7-k}{8}\right)\,\frac{1}{T},
\nonumber\\
&\Rightarrow\;F(T) = F_0\,T^{\frac{7-k}{8}} ,
\end{align}
where $F_0$ is an integration constant and $k \neq 7$. By comparison with ref \cite{SSpaper} solution, we can set $A_2=A_3^{\frac{1+k}{4}}$ leading to $c_0=1$ for all $k$ and $c_0=-1$ for $k=8k'-1$ where $k'$ is an integer. Eqn \eqref{354} is a power-law $F(T)$ solution as in ref \cite{SSpaper}, but the $A_2=A_3^n$ ansatz also leads to a similar power-law solution as shown in \cite{SSpaper}.

\subsection{$A_3=c_0\,t$ solutions}\label{sect32}

In section \ref{sect31}, we find that $A_3=c_0\,t$ is the only possible solution for $A_3$. From this point, we may find some possible $A_2$ exact expression and $F(T)$ solutions which are not necessarily a power-law solution. The eqns \eqref{351d} to \eqref{351c} become:
\begin{subequations}
\begin{align}
T &=  \frac{4}{t} \left(\ln(A_2)\right)'+\frac{2}{t^2} - \frac{2}{c_0^2\,t^2} , \label{360d}
\\
B' &= \frac{\left( -\left(\ln(A_2)\right)'\,t^{-1} -\left(\ln(A_2)\right)''-\left(\ln(A_2)\right)'^2+ \frac{\left(c_0^{-2}+1\right)}{t^2} \right)}{\left(\ln(A_2)\right)' - t^{-1}} , \label{360a}
\\
F(T) &= 2\left( T + \frac{2}{c_0^2\,t^2} \right) F_T(T), \label{360b}
\\
B' &=  \left(\ln(A_2)\right)' . \label{360c}
\end{align}
\end{subequations}
By substituting eqn \eqref{360c} into \eqref{360a}, we obtain the DE for $A_2$:
\begin{align}
0=\left(\ln(A_2)\right)''+2\left(\ln(A_2)\right)'^2 - \frac{1+c_0^{-2}}{t^2} . \label{360e}
\end{align}
By setting $y=\ln(A_2)$ and $c_0^{-2}=\frac{1}{8}(k-3)(k+3)$, eqn \eqref{360e} becomes $y''(t)+2y'(t)^2+\left(\frac{1-k^2}{8}\right)\,t^{-2}=0$. The general solution is exactly:
\begin{align}
y=\ln(A_2)=\frac{1}{2}\ln\left[t^{k}+y_1\right]+\frac{1}{4}\left(1-k\right)\,\ln(t) , \label{360f}
\end{align}
which leads to:
\begin{align}\label{360g}
A_2=\left[t^{k}+y_1\right]^{\frac{1}{2}}\,t^{\frac{1}{4}\left(1-k\right)} ,
\end{align}
where $y_1$ is an arbitrary constant. Then by using eqns \eqref{360f} and \eqref{360g}, eqn \eqref{360d} becomes:
\begin{align}
T &=  \frac{2k\,t^{-2}}{\left[1+y_1\,t^{-k}\right]}-\frac{(k-3)(k+7)}{4}\,t^{-2} . \label{360h}
\end{align}

\noindent ${\bf y_1=0}$ case: We obtain that eqn \eqref{360h} is the same as eqn \eqref{353}. Then by substituting this eqn \eqref{353} into eqn \eqref{360b}, we refind exactly the eqn \eqref{354}.

\noindent ${\bf y_1\neq 0}$ case: Eqn \eqref{360h} will be a characteristic eqn. for $t(T)$ relationship before solving eqn \eqref{360b} for $F(T)$ specific solution. For this, we set in eqn \eqref{360h} the parameter $k$ and then eqn \eqref{360h} becomes:
\begin{align}
T &=  \frac{2k\,t^{-2}}{\left[1+y_1\,t^{-k}\right]}-\left(\frac{(k-3)(k+7)}{4}\right)\,t^{-2} , \label{362a}
\end{align}
where $k>3$ and $k<-3$ for a real value of $c_0$. Then eqn \eqref{360b} will be simplified as:
\begin{align}\label{362b}
F(T) &= 2\left( T + \frac{(k-3)(k+3)}{4}\,t^{-2}(T) \right) F_T(T)
\nonumber\\
\Rightarrow\,& F(T)=F(0)\, \exp\Bigg[\frac{1}{2}\int\,\frac{dT}{\left( T + \frac{(k-3)(k+3)}{4} t^{-2}(T)\right)}\Bigg]
\end{align}
There are some specific value of $k$ (we set integer values of $k$ for eqn \eqref{362a} possible exact solutions) leading to analytic $t(T)$ solutions for eqn \eqref{362a}:
\begin{enumerate}

\item \textbf{Limit of} ${\bf k=\pm 3 }$: If $c_0\,\rightarrow\,\pm\infty$, we obtain that eqn \eqref{362b} simplifies as:
\begin{align}\label{362c}
F(T) &= 2T\,F_T(T)
\nonumber\\
\Rightarrow\, & F(T)=F_0\,\sqrt{T}.
\end{align}
We find as limit an usual power-law solution for $k\,\rightarrow\,\pm 3$. {  This eqn \eqref{362c} is also obtained for the case of dark energy  perfect fluid where the EoS is $P=-\rho$ in ref \cite{Amir2015}.}

\item ${\bf k=\pm 4 }$ and ${\bf \pm 6 }$ cases: We obtain that $c_0^2=\frac{8}{7}$ and $\frac{8}{27}$ respectively. We find respectively degree $3$ and $4$ characteristic equations from eqn \eqref{362a} and then eqn \eqref{362b} will not lead in both cases to an analytical and closed form for $F(T)$.

\end{enumerate}
All these subcases lead to new teleparallel $F(T)$ solutions and confirm some recent solutions such as those in ref \cite{Amir2015}.

\subsection{Exponential ansatz solutions}

We can also introduce another approach for solution by using an exponential ansatz. We can see this ansatz as an infinite superposition of power-law terms as:
\begin{align}\label{380}
A_2(t)= \,\exp(b\,t) =\sum_{n=0}^{\infty}\,\frac{(b\,t)^n}{n!},\quad\quad A_3(t)=c_0\,\exp(c\,t) =c_0\,\sum_{n=0}^{\infty}\,\frac{(c\,t)^n}{n!}.
\end{align}
By substituting eqns \eqref{380}, eqns \eqref{351d} to \eqref{351c} become:
\begin{subequations}
\begin{align}
T &= 2c(c+2b) - \frac{2}{c_0^2}\,\exp(-2c\,t) , \label{381d}
\\
B' &= -(2c+b)+ \frac{1}{c_0^2(b-c)}\,\exp(-2c\,t), \label{381a}
\\
F(T) &= 2\left( T + \frac{2}{c_0^2}\,\exp(-2c\,t) \right) F_T(T), \label{381b}
\\
0 &= c\left[B' + c - b \right] . \label{381c}
\end{align}
\end{subequations}
From eqn \eqref{381c}, we find two GR solutions:
\begin{itemize}
\item $c=0$ case: We obtain from eqns \eqref{381d} to \eqref{381b} that $T=- \frac{2}{c_0^2}=$ constant, $A_3=c_0=$ constant, $B'=-b+\frac{1}{c_0^2\,b}$ and $F(T)=0$. 

\item $B'=b-c$: Eqn \eqref{381a} leads to $c=0$, then $A_3=c_0$ constant and $b=\frac{\delta_1}{\sqrt{2}\,c_0}$ leading to $A_2= \,\exp\left(\frac{\delta_1\,t}{\sqrt{2}\,c_0}\right)$ for $t$-independent solution. Then eqn \eqref{381d} leads to $T=- \frac{2}{c_0^2}=$ constant and finally eqn \eqref{381b} leads to $F(T)=0$. 
\end{itemize}
No purely teleparallel $F(T)$ solution is possible for exponential ansatz in vacuum.


\section{Linear perfect fluid solutions}\label{sect4}

A perfect isotropic fluid with a linear EoS $P=\alpha\,\rho$, where $-1<\alpha \leq 1$, is now the matter source. We solve eqn \eqref{302d} for a specific $\rho(t)$ expression in terms of $A_2$ and $A_3$:
\begin{align}
&\left(1+\alpha\right)\,\left(\ln(A_2\,A_3^2)\right)'+\left(\ln\,\rho\right)'=0,
\nonumber\\
&\quad\Rightarrow\;\rho(t)=\frac{\rho_0}{\left(A_2(t)\,A_3^2(t)\right)^{\left(1+\alpha\right)}} \label{401a}
\end{align}
The eqns \eqref{301a} to \eqref{301c} become:
\begin{subequations}
\begin{align}
 T =& 2\left(\ln(A_3)\right)'\left(\ln\left(A_2^2\,A_3\right)\right)' - \frac{2}{A_3^2}  , \label{401e}
\\
 B' =&  -\left(\ln\left(A_2\,A_3^2\right)\right)'+\frac{\frac{1}{A_3^2} - \left(\ln\left(\frac{A_2}{A_3}\right)\right)''}{\left(\ln\left(\frac{A_2}{A_3}\right)\right)' }, \label{401d} 
\\
 \kappa \rho + \frac{F(T)}{2} =& \left( T + \frac{2}{A_3^2} \right) F_T(T), \label{401b}
\\
 -\kappa \rho =& \frac{F_T(T)}{\left(1+\alpha\right)} \left[\left(\ln(A_3)\right)'\left(B'+ \left(\ln(A_3)\right)'-\left(\ln(A_2)\right)'\right)+ \left(\ln(A_3)\right)''\right] . \label{401c}
\end{align}
\end{subequations}
By adding eqns \eqref{401b} and \eqref{401c} and then substituting eqn \eqref{401d} , we obtain a linear DE in $F(T)$:
\small
\begin{align}
F(T) &= \frac{2\,F_T(T)}{\left(1+\alpha\right)} \Bigg[\left(\alpha+\frac{1}{2}\right)\left(T + \frac{2}{A_3^2}\right)+\frac{\left(\ln(A_3)\right)'\left(\frac{1}{A_3^2} - \left(\ln\left(A_2\right)\right)''\right)+\left(\ln\left(A_2\right)\right)'\left(\ln(A_3)\right)''}{\left(\ln\left(\frac{A_2}{A_3}\right)\right)' } \Bigg] . \label{402}
\end{align}
\normalsize

\subsection{Power-law solutions}\label{sect41}

As in section \ref{sect31}, we use eqn \eqref{350} ansatz. Then eqns \eqref{401a} to \eqref{401c} become:
\begin{subequations}
\begin{align}
T &= \frac{2c(c+2b)}{t^2} - \frac{2}{c_0^2\,t^{2c}}, \label{420a}
\\
B' &= \left[\frac{(1-b-2c)}{t}+\frac{t^{1-2c}}{c_0^2\,(b-c)}\right], \label{420b}
\\
\kappa \rho + \frac{F(T)}{2} &= \left( T + \frac{2}{c_0^2}\,t^{-2c} \right) F_T(T), \label{420c}
\\
-\kappa \rho &=\frac{F_T(T)\,c}{\left(1+\alpha\right)} \left[ \frac{1}{t} B' + \frac{(c-b-1)}{t^2} \right], \label{420d}
\\
\rho&=\rho_0\,\left(c_0^2\right)^{-\left(1+\alpha\right)}\,t^{-\left(1+\alpha\right)(b+2c)}. \label{420e}
\end{align}
\end{subequations}
With eqn \eqref{420a}, we find the characteristic eqn to solve for each specific value of $c$:
\begin{align}\label{422}
0=2c(c+2b)\,t^{-2} - \frac{2}{c_0^2}\,t^{-2c}-T.
\end{align}
By putting eqn \eqref{420c} and \eqref{420d} together and then substituting eqns \eqref{420a} and \eqref{420b}, we obtain as equation:
\begin{align}
F(T) &= F_T(T)\Bigg[\frac{\left(1+2\alpha\right)}{\left(1+\alpha\right)}\left(T+\frac{2}{c_0^2}\,t^{-2c}(T)\right)+\frac{2c\,t^{-2c}(T)}{c_0^2\,\left(1+\alpha\right)(b-c)}\Bigg], 
\nonumber\\
\Rightarrow\,& F(T)=F(0)\,\exp\Bigg[(1+\alpha)\int\,dT\,\Bigg[(1+2\alpha)T+\frac{2}{c_0^2}\left(1+2\alpha+\frac{c}{(b-c)}\right)\,t^{-2c}(T)\Bigg]^{-1}\Bigg].\label{421}
\end{align}

\noindent For ${\alpha=-\frac{1}{2}}$ cases, eqn \eqref{421} becomes more simple as:
\begin{align}
 F(T)=F(0)\,\exp\Bigg[\frac{c_0^2(b-c)}{4c}\,\int\,dT\,t^{2c}(T)\Bigg], \label{421s}
\end{align}
where $b \neq c$.

\noindent For the particular ${\bf c=-2b}$ case, eqn \eqref{422} simplifies as:
\begin{align}\label{422a}
t^{4b}(T)=\frac{c_0^2}{2}\,(-T).
\end{align}
Then eqn \eqref{421} becomes:
\begin{align}
F(T)=& F(0)\,T^{\frac{3\left(1+\alpha\right)}{2}} .\label{421a}
\end{align}
The eqn \eqref{421a} is a pure power-law solution for $\alpha \neq -1$.

\noindent There are in principle solutions for eqn \eqref{422} in the cases ${ \bf c=\frac{1}{2},\,-\frac{1}{2},\,1,\,-1,\,\frac{3}{2},\,2,\,-2,\,3,}$ ${  \bf -3,\,4 }$ ($c \neq -2b$ and $c \neq 0$). For analytically solvable solutions, we will solve for the following cases:
\begin{enumerate}
\item ${\bf c=\frac{1}{2}}$: Eqn \eqref{422} becomes:
\begin{align}\label{431}
& 0=\left(\frac{1}{2}+2b\right)\,t^{-2} - \frac{2}{c_0^2}\,t^{-1}-T,
\nonumber\\
&\Rightarrow\,t^{-1}(T)=\frac{\left[1+\delta_1\,\sqrt{1+\left(\frac{1}{2}+2b\right)\,c_0^4\,T}\right]}{\left(\frac{1}{2}+2b\right)\,c_0^2}
\end{align}
Eqn \eqref{421} becomes by substituting eqn \eqref{431} (i.e. $b \neq -\frac{1}{4}$ and $\alpha \neq -\frac{1}{2}$):
\small
\begin{align}
F(T)=& F(0)\,\Bigg[-\left(\frac{1}{2}+2b\right)\,c_0^4\Bigg((1+2\alpha)\,T+\frac{4}{c_0^4}\left(1+2\alpha+\frac{1}{(2b-1)}\right)
\nonumber\\
& \times\,\frac{\left[1+\delta_1\,\sqrt{1+\left(\frac{1}{2}+2b\right)\,c_0^4\,T}\right]}{\left(1+4b\right)}\Bigg)\Bigg]^{\frac{\left(1+\alpha\right)}{\left(1+2\alpha\right)}} \,\exp\Bigg[2(1+\alpha)\left(2b-1+\frac{1}{1+2\alpha}\right)
\nonumber\\
& \times\,\tanh^{-1}\left[\left(1+2\alpha\right)(2b-1)\left(1+\delta_1\sqrt{1+2 c_0^4\left(\frac{1}{4}+b\right)\,T}\right)+1\right]\Bigg], \label{432}
\end{align}
\normalsize
where $b \neq \left\lbrace-\frac{1}{4},\,\frac{1}{2}\right\rbrace$, $\alpha \neq \left\lbrace -1,\,-\frac{1}{2}\right\rbrace$ and $\delta_1=\pm 1$. For $\alpha=-\frac{1}{2}$, eqn \eqref{421s} becomes by using eqn \eqref{431}:
\small
\begin{align}
F(T)=F(0)\,\left[1+\delta_1\,\sqrt{1+\left(\frac{1}{2}+2b\right)\,c_0^4\,T}\right]^{\frac{1}{2}-b}\,\exp\Bigg[\delta_1\,\left(b-\frac{1}{2}\right)\,\sqrt{1+\left(\frac{1}{2}+2b\right)\,c_0^4\,T}\Bigg]   ,\label{432s}
\end{align}
\normalsize
where $b \neq \frac{1}{2}$.

\item ${\bf c=1}$: Eqn \eqref{422} becomes:
\begin{align}\label{435}
& 0=\left(2(1+2b) - \frac{2}{c_0^2}\right)\,t^{-2}-T,
\nonumber\\
&\Rightarrow\,t^{-2}(T)=\frac{T}{2\left(1+2b - \frac{1}{c_0^2}\right)}.
\end{align}
Then eqn \eqref{421} becomes with eqn \eqref{435} (i.e. $\alpha \neq -\frac{1}{2}$):
\begin{align}\label{436}
F(T)=F(0)\,T^{\frac{(1+\alpha)(b-1)\left(c_0^2(1+2b) - 1\right)}{c_0^2(1+2\alpha)(2b+1)(b-1)+1}},
\end{align}
where $b \neq 1$. We obtain a power-law $F(T)$ solution and eqn \eqref{436} is the most general solution for the linear $A_3=c_0\,t$ subcase. For $\alpha=-\frac{1}{2}$, eqn \eqref{421s} becomes with eqn \eqref{435}:
\begin{align}
F(T)=F(0)\,T^{\frac{(b-1)}{2}\,\left[(1+2b)\,c_0^2-1\right]}   , \label{436s}
\end{align}
where $b \neq 1$.

\item ${\bf c=-1}$: Eqn \eqref{422} becomes:
\begin{align}\label{437}
& 0=t^{4}+\frac{c_0^2\,T}{2}\,t^{2} - c_0^2(1-2b), \quad &
\nonumber\\
&\Rightarrow\,t^{2}(T)= \frac{c_0^2}{4}\left[-T+\delta_1\,\sqrt{T^2+16(1-2b)\,c_0^{-2}}\right] .
\end{align}
Then eqn \eqref{421} becomes with eqn \eqref{437} {  (i.e. $\alpha \neq -\frac{1}{2}$)}:
\begin{align}
F(T)=& \,F(0)\,\left[\frac{\left[\frac{4(1+2\alpha)}{(b+1)}\,T^2-16\,\left(1+2\alpha-\frac{1}{(b+1)}\right)^2\,\frac{(1-2b)}{c_0^2}\right]^{\left(b+\frac{2(1+\alpha)}{(1+2\alpha)}\right)}}{\left[T+\sqrt{T^2+16(1-2b)\,c_0^{-2}}\right]^{2\delta_1\left(b+\frac{2\alpha}{(1+2\alpha)}\right)}}\right]^{\frac{(1+\alpha)}{4}}
\nonumber\\
&\quad\times\, \exp\Bigg[\frac{(1+\alpha)}{2}\left(b+\frac{2(1+\alpha)}{(1+2\alpha)}\right)
\nonumber\\
&\quad\times\, \tanh^{-1}\left[\frac{\delta_1\,\left(1+2\alpha+\frac{1}{(b+1)}\right)\,T}{\left(1+2\alpha-\frac{1}{(b+1)}\right)\,\sqrt{T^2+16(1-2b)\,c_0^{-2}}}\right]\Bigg] , \label{438}
\end{align}
where $b \neq -1$, $\alpha \neq \left\lbrace -1,\,-\frac{1}{2}\right\rbrace$  and $\delta_1=\pm 1$. For $\alpha=-\frac{1}{2}$, eqn \eqref{421s} becomes with eqn \eqref{437}:
\small
\begin{align}
F(T)=& F(0)\,\left[T+\sqrt{T^2+16(1-2b)\,c_0^{-2}}\right]^{-\frac{\delta_1}{2}(b+1)}
\nonumber\\
&\,\times\,\exp\Bigg[-\frac{c_0^2\,(b+1)}{32(1-2b)}\,T\left(T+\delta_1\,\sqrt{T^2+16(1-2b)\,c_0^{-2}}\right)\Bigg]   ,\label{438s}
\end{align}
\normalsize
where $b \neq \left\lbrace -1,\,\frac{1}{2}\right\rbrace$.

\item ${\bf c=2}$: Eqn \eqref{422} becomes:
\begin{align}\label{441}
& 0=t^{-4}-4c_0^2\,(1+b)\,t^{-2} +\frac{c_0^2\,T}{2},
\nonumber\\
& \Rightarrow\,t^{-2}(T)=2c_0^2\left[(1+b)+\delta_1\,\sqrt{(1+b)^2-\frac{T}{8c_0^2}}\right],
\end{align}
where $\delta_1=\pm 1$. Then eqn \eqref{421} becomes with eqn \eqref{441} {  (i.e. $\alpha \neq -\frac{1}{2}$)}:
\begin{align}
& F(T) =\,F(0)\,\exp\Bigg[(1+\alpha)\left(2-b-\frac{2}{1+2\alpha} \right)
\nonumber\\
&\;\times\,\tanh^{-1}\left[1+\frac{2}{(1+2\alpha)(b-2)}\left(1+\delta_1\,\sqrt{1-\frac{T}{8c_0^2(1+b)^2}}\right)\right]\Bigg]
\nonumber\\
&\;\times\,\Bigg[\frac{2\,T}{2-b}+16\,c_0^2(1+b)^2\left(1+2\alpha+\frac{2}{(b-2)}\right)\left(1+\delta_1\,\sqrt{1-\frac{T}{8c_0^2(1+b)^2}}\right)\Bigg]^{\frac{(1+\alpha)(2-b)}{2}}, \label{442}
\end{align}
where $b \neq 2$, $\alpha \neq \left\lbrace -1,\,-\frac{1}{2}\right\rbrace$ and $\delta_1=\pm 1$. For $\alpha=-\frac{1}{2}$, eqn \eqref{421s} becomes with eqn \eqref{441}:
\small
\begin{align}
F(T)=F(0)\,\left[4(1+b)+4\delta_1\,\sqrt{(1+b)^2-\frac{T}{8c_0^2}}\right]^{\frac{(2-b)}{2}}\,\exp\Bigg[\frac{(2-b)(1+b)}{2\left((1+b)+\delta_1\,\sqrt{(1+b)^2-\frac{T}{8c_0^2}}\right)}\Bigg]   ,\label{442s}
\end{align}
\normalsize
where $b \neq 2$.

\end{enumerate}

For the other values of $c$, their eqn \eqref{421} integral will not lead to analytic and closed $F(T)$ solutions. All these previous teleparallel $F(T)$ solutions are new results. However, the eqns \eqref{436} and \eqref{436s} for the case $c=1$ are a bit similar to the solutions of ref. \cite{Amir2015}. In this last paper, the solutions obtained are only for $\alpha=-1$, $0$, $\frac{1}{3}$ and combinations of these solutions.


\subsection{$A_2=A_3^n$ ansatz solutions}\label{sect42}

There are other possible solutions of the eqns \eqref{401a} to \eqref{401c} by using the ansatz $A_2=A_3^n$ as in ref \cite{SSpaper}. In this case, eqns \eqref{401a} to \eqref{401c} become:
\begin{subequations}
\begin{align}
T =& 2(1+2n)\left(\ln(A_3)\right)'^2 - \frac{2}{A_3^2}, \label{403a}
\\
B' =&  -\frac{\left((n+2)\left(\ln(A_3)\right)'^2 +  \left(\ln(A_3)\right)'' + \frac{1}{(n-1)\,A_3^2} \right)}{\left(\ln(A_3)\right)'} ,  \label{403b}
\\
\kappa \rho + \frac{F(T)}{2} =& \left( T + \frac{2}{A_3^2} \right) F_T(T), \label{403d}
\\
-\kappa \rho =&\frac{F_T(T)}{\left(1+\alpha\right)} \left[ \left(\ln(A_3)\right)' B' + \left(\ln(A_3)\right)''- (n-1)\,\left(\ln(A_3)\right)'^2 \right] , \label{403e}
\\
\rho=&\rho_0\,A_3^{-(2+n)\left(1+\alpha\right)}, \label{403c}
\end{align}
\end{subequations}
where $n \in \mathbb{R}$. We find from eqn \eqref{403a} that:
\begin{align}\label{403aa}
(2n+1)\left(\ln(A_3)\right)'^2=\frac{1}{2}\left(T+\frac{2}{A_3^2}\right).
\end{align}
By putting eqns \eqref{403b} to \eqref{403e} together, and then by substituting eqn \eqref{403aa}, we find that
\begin{align}
F(T) &= \,F_T(T)\Bigg[\left(\frac{1+2\alpha}{1+\alpha}\right)\,T + \frac{2\left(n\left(1+2\alpha\right)-2(1+\alpha)\right)}{A_3^2(n-1)\left(1+\alpha\right)}\Bigg]. \label{404} 
\end{align}
The eqn \eqref{404} solutions depend on the $A_3$ components. A constant $A_3$ leads to constant torsion scalar according to eqn \eqref{403a} and then to GR solution. So, we need that $A_3 \neq$ constant and then a $F(T)$ solution not depending on $A_3$. For this requirement, we need to satisfy the relation $n\left(1+2\alpha\right)-2(1+\alpha)=0$ leading to the solution $n=\frac{2(1+\alpha)}{1+2\alpha}$ where $\alpha \neq -\frac{1}{2}$ for a power-law solution. The eqn \eqref{404} leads to the following $F(T)$ solutions:
\begin{enumerate}
\item ${  \alpha \neq -\frac{1}{2}}$: Eqn \eqref{404} becomes a simple DE:
\begin{align}
F(T) &= \,\left(\frac{1+2\alpha}{1+\alpha}\right)\,T\,F_T(T),
\nonumber\\
\Rightarrow\,&\,F(T) = \,F_0\,T^{\frac{1+\alpha}{1+2\alpha}} \equiv\,F_0\,T^{\frac{n}{2}}, \label{406} 
\end{align}
where $F_0$ is an integration constant, $\alpha \neq \left\lbrace-\frac{1}{2},\,-1,\,0\right\rbrace$ and $A_2=A_3^{\frac{2(1+\alpha)}{1+2\alpha}}$. We also refind the eqn \eqref{406} by the general way. The dust matter case $\alpha=0$ leads to GR solutions.

\item ${\alpha = -\frac{1}{2}}$: We can approximate eqn \eqref{404} by setting $\alpha=-\frac{1}{2}+\Delta\alpha$ for very small $\Delta\alpha$ as:
\begin{align}
F(T) &\approx \,4\,F_T(T)\Bigg[\Delta\alpha\,T - \frac{\left(1-2n\Delta\alpha\right)}{A_3^2(n-1)}\Bigg]. \label{404a} 
\end{align}
If $n\,\rightarrow\,\infty$ and $\Delta\alpha \,\rightarrow\,0$ for all $A_3$, we obtain the GR solutions. For an $A_3$ independent $F(T)$ solution, we need to satisfy the condition $\Delta\alpha \approx \frac{1}{2n}$ for a large $n$. In this case, eqn \eqref{404a} becomes for $n\gg 1$:
\begin{align}
F(T) &\approx \,\frac{2}{n}\,T\,F_T(T). \label{404b} 
\end{align}
The solution is exactly eqn \eqref{406} and then eqn \eqref{404b} proves that function $F(T)$ is defined around $\alpha=-\frac{1}{2}$ and $n$ is very large.

\end{enumerate}

We need to find possible $A_3$ satisfying to eqns \eqref{403a} to \eqref{403c} for eqn \eqref{406} solution for all $\alpha\neq -\frac{1}{2}$ cases. By using eqns \eqref{403d} and \eqref{403c} and then by substituting eqn \eqref{406}, we obtain as characteristic eqn. for $T$:
\begin{align}
0 &= \frac{2\kappa \rho_0}{F_0}\,A_3^{-\frac{2(1+\alpha)(2+3\alpha)}{1+2\alpha}} -\left(\frac{1}{1+2\alpha}\right)\,T^{\frac{1+\alpha}{1+2\alpha}} - \frac{4}{A_3^2}\left(\frac{1+\alpha}{1+2\alpha}\right) \,T^{-\frac{\alpha}{1+2\alpha}}. \label{407}
\end{align}
Then eqn \eqref{403a} in terms of $\alpha$ is:
\begin{align}\label{408}
T =& 2\left(\frac{5+6\alpha}{1+2\alpha}\right)\frac{A_3'^2}{A_3^2} - \frac{2}{A_3^2}.
\end{align}
From eqn \eqref{407}, we can isolate $T$ in term of $A_3$ and then equate to eqn \eqref{408} leading to the DE for $A_3(t)$ depending on $\alpha$ and power of $T$ inside eqn \eqref{407}. There are possible simple analytical solutions:
\begin{enumerate}
\item ${\alpha=-\frac{1}{3}}$ case: Eqn \eqref{406} becomes as simple as $F(T) = \,F_0\,T^{2}$, $n=4$ and eqn \eqref{403c} leads to $\rho=\rho_0 A_3^{-4}$. Eqn \eqref{407} and \eqref{408} become:
\begin{subequations}
\begin{align}
0 =& T^{2}+\frac{8}{3A_3^2} \,T -\frac{2\kappa \rho_0}{3F_0}\,A_3^{-4}, \label{409a}
\\
T =& 18\frac{A_3'^2}{A_3^2} - \frac{2}{A_3^2}. \label{409b}
\end{align}
\end{subequations}
By putting eqn \eqref{409a} and \eqref{409b} together, we find as DE and solution:
\begin{align}
A_3'=&\delta_2\,\sqrt{\frac{2}{27}\left(\frac{1}{2}+\delta_1\,\sqrt{1+\frac{3\kappa \rho_0}{8F_0}}\right)} =\text{constant} , 
\nonumber\\
& \Rightarrow\,A_3=\delta_2\,\sqrt{\frac{2}{27}\left(\frac{1}{2}+\delta_1\,\sqrt{1+\frac{3\kappa \rho_0}{8F_0}}\right)}\,t =c_0\,t \label{410}
\end{align}
where $\left(\delta_1,\,\delta_2\right)=\left(\pm 1,\,\pm 1\right)$. Eqn \eqref{410} leads to a linear $A_3$ and this confirms the power-law ansatz result.

\item ${\alpha =-\frac{2}{3}}$ case: Eqn \eqref{406} becomes $F(T) = \,F_0\,T^{-1}$, $n=-2$ and eqn \eqref{403c} leads to $\rho(t)=\rho_0=$ constant. Then eqns \eqref{407} and \eqref{408} become:
\begin{subequations}
\begin{align}
0 &= \frac{2\kappa \rho_0}{F_0} +3\,T^{-1} + \frac{4}{A_3^2}\,T^{-2},  \label{410aa}
\\
T =& -6\frac{A_3'^2}{A_3^2} - \frac{2}{A_3^2}. \label{410ab}
\end{align}
\end{subequations}
By putting eqns \eqref{410aa} and \eqref{410ab} together, we obtain as DE:
\begin{align}
3A_3'^2=&\frac{3F_0}{8\kappa\,\rho_0}\,A_3\,\left[A_3+\delta_1\,\sqrt{A_3^2-\frac{32\kappa\,\rho_0}{9F_0}}\right]-1  ,
\nonumber\\
\Rightarrow\,\frac{\delta_2\,(t-t_0)}{\sqrt{3}} =& \int\,dA_3\,\Bigg[\frac{3F_0}{8\kappa\,\rho_0}\,A_3\,\left[A_3+\delta_1\,\sqrt{A_3^2-\frac{32\kappa\,\rho_0}{9F_0}}\right]-1\Bigg]^{-1} ,  \label{410ac}
\end{align}
where $\delta_2=\pm 1$. This last integral is complex to solve. However, there are two possible limits:
\begin{enumerate}
\item Low fluid density limit $\frac{\kappa\,\rho_0}{F_0} \ll 1$: Eqn \eqref{410ac} simplifies for $\delta_1=+1$:
\begin{align}
\delta_2\,(t-t_0) \approx & -\frac{4\kappa\,\rho_0}{\sqrt{3}F_0}\,A_3^{-1} ,
\nonumber\\
\Rightarrow\,A_3 \approx & \frac{4\delta_2\,\kappa\,\rho_0}{\sqrt{3}F_0\,(t_0-t)} .\label{410ad}
\end{align}
We find at eqn \eqref{410ad} a $A_3$ component for a contracting universe.

\item High fluid density limit $\frac{\kappa\,\rho_0}{F_0} \gg 1$: Eqn \eqref{410ac} simplifies for $\delta_1=+1$:
\begin{align}
\frac{\delta_2\,(t-t_0)}{\sqrt{3}} \approx & -A_3+\delta_1\sqrt{-\frac{F_0}{8\kappa\,\rho_0}}\,A_3^2 ,
\nonumber\\
 \Rightarrow\,A_3 \approx & \left(-\frac{2\kappa\,\rho_0}{F_0}\right)^{1/2} \left[1+\delta_3\sqrt{1+\frac{2\delta_1\,\delta_2}{\sqrt{3}} \left(-\frac{F_0}{2\kappa\,\rho_0}\right)^{1/2} (t-t_0)}\right],
\nonumber\\ 
 \approx & \delta_1 \left(-\frac{2\kappa\,\rho_0}{F_0}\right)^{1/2}\left(1+\delta_3\right) +\frac{\delta_2\,\delta_3 }{\sqrt{3}}\,(t-t_0), \label{410ae}
\end{align}
where $\delta_3=\pm 1$. We find from eqn \eqref{410ae} a linear $A_3$ as the highest fluid density limit as for vacuum solutions and also confirms the power-law ansatz result.
\end{enumerate}

\item ${\alpha =-\frac{1}{4}}$ case: Eqn \eqref{406} becomes $F(T) = \,F_0\,T^{\frac{3}{2}}$, $n=3$ and eqn \eqref{403c} leads to $\rho(t)=\rho_0\,A_3^{-15/4}$. Then eqns \eqref{407} and \eqref{408} become:
\begin{subequations}
\begin{align}
0 &= T^{\frac{3}{2}} + \frac{3}{A_3^2} \,T^{\frac{1}{2}}-\frac{\kappa \rho_0}{F_0}\,A_3^{-\frac{15}{4}},  \label{411a}
\\
T =& 14\,\left(\ln(A_3)\right)'^2 - \frac{2}{A_3^2}.  \label{411b}
\end{align}
\end{subequations}
By putting together eqns \eqref{411a} and \eqref{411b}, we find the DE:
\small 
\begin{align}\label{412}
A_3'=& \frac{\delta_1}{\sqrt{14}\,2^{1/3}\,A_3^{1/4}}\Bigg[\Bigg(\sqrt{\frac{\left(\kappa\,\rho_0\right)^2}{F_0^2}+4\,A_3^{3/2}}+\frac{\kappa\,\rho_0}{F_0}\Bigg)^{2/3}
\nonumber\\
&\,+ 2^{4/3}\,A_3\Bigg(\sqrt{\frac{\left(\kappa\,\rho_0\right)}{F_0^2}+4\,A_3^{3/2}}+\frac{\kappa\,\rho_0}{F_0}\Bigg)^{-2/3}\Bigg]^{1/2}
\nonumber\\
\Rightarrow\,& \frac{\delta_1\,t}{\sqrt{14}} = \int\,dA_3\,\Bigg[\Bigg(\sqrt{1+\left(\frac{\kappa\,\rho_0}{2\,F_0\,A_3^{3/4}}\right)^2}+\frac{\kappa\,\rho_0}{2\,F_0\,A_3^{3/4}}\Bigg)^{2/3}
\nonumber\\
&\quad\quad\quad\quad+\Bigg(\sqrt{1+\left(\frac{\kappa\,\rho_0}{2\,F_0\,A_3^{3/4}}\right)^2}+\frac{\kappa\,\rho_0}{2\,F_0\,A_3^{3/4}}\Bigg)^{-2/3}\Bigg]^{-1/2}
\end{align}
\normalsize
Eqn \eqref{412} is complex to solve. However, there are two limit cases where we can solve this equation:
\begin{enumerate}
\item  Low fluid density limit $\frac{\kappa\,\rho_0}{2\,F_0\,A_3^{3/4}}\ll 1$: In this situation, eqn \eqref{412} will be approximated at the $1$st order level:
\begin{align}\label{413}
\frac{\delta_1\,t}{\sqrt{7}} \approx & \int\,dA_3\,\left[1-\frac{1}{36}\,\left(\frac{\kappa\,\rho_0}{F_0}\right)^2\,A_3^{-3/2}\right],
\nonumber\\
\approx & A_3+\frac{1}{18\,\sqrt{A_3}}\,\left(\frac{\kappa\,\rho_0}{F_0}\right)^2+C_1,
\nonumber\\
\Rightarrow\,&\,A_3^{3/2}-\frac{\delta_1\,\left(t-t_0\right)}{\sqrt{7}}\,A_3^{1/2}+\frac{1}{18}\,\left(\frac{\kappa\,\rho_0}{F_0}\right)^2 \approx 0 ,
\end{align}
where $C_1$ is an integration constant and $t_0$ is depending on this constant. For weak $\frac{\kappa\,\rho_0}{F_0}$, the only relevant solution for eqn \eqref{413} leading to a real function for $A_3$ is with $\delta_1=-1$ subcase:
\small
\begin{align}\label{414}
A_3(t) \approx &\frac{1}{6^{4/3}\cdot 7^{2/3}}\Bigg(\sqrt{48\sqrt{7}\,\left(t-t_0\right)^3+49\left(\frac{\kappa\,\rho_0}{F_0}\right)^4}-7\left(\frac{\kappa\,\rho_0}{F_0}\right)^2\Bigg)^{2/3} 
\nonumber\\
&\;+\frac{4\,\left(t-t_0\right)^2}{6^{2/3}\,\sqrt[3]{7}}\Bigg(\sqrt{48\sqrt{7}\,\left(t-t_0\right)^3+49\left(\frac{\kappa\,\rho_0}{F_0}\right)^4}-7\left(\frac{\kappa\,\rho_0}{F_0}\right)^2\Bigg)^{-2/3} - \frac{2\,\left(t-t_0\right)}{3\sqrt{7}}
\end{align}
\normalsize
For $\frac{\kappa\,\rho_0}{F_0}$ very weak limit, eqn \eqref{414} will simplify as:
\begin{align}\label{414a}
A_3(t) \approx & \frac{7}{324}\,\left(\frac{\kappa\,\rho_0}{F_0}\right)^4\,\left(t-t_0\right)^{-2},
\nonumber\\
\approx & c_0\,t^{-2} \quad\quad\quad\quad \text{for }\,t_0=0.
\end{align}

\item High fluid density limit $\frac{\kappa\,\rho_0}{2\,F_0\,A_3^{3/4}}\gg 1$ (or $\frac{F_0}{\kappa\,\rho_0}\ll 1$): In this last case, eqn \eqref{412} will be approximated as:
\small
\begin{align}\label{415}
\frac{\delta_1\,t}{\sqrt{14}} \approx & \left(\frac{F_0}{\kappa\,\rho_0}\right)^{1/3}\,\int\,dA_3\,A_3^{1/4},
\nonumber\\
\approx & \frac{4}{5}\left(\frac{F_0}{\kappa\,\rho_0}\right)^{1/3}\,A_3^{5/4}+C_2,
\nonumber\\
\rightarrow\,A_3(t)\,\approx &\,\left(\frac{5}{4\sqrt{14}}\right)^{4/5}\left(\frac{F_0}{\kappa\,\rho_0}\right)^{4/15}\,\left(t-t_0\right)^{4/5},
\nonumber\\
\approx & \,c_0\,t^{4/5} \quad\quad\quad\quad \text{for }\,t_0=0.
\end{align}
\normalsize
where $C_2$ is an integration constant and $t_0$ is depending on this constant.

\end{enumerate}

\end{enumerate}
All these previous teleparallel $F(T)$ solutions are new results. However, there are some specific solutions similar to the cases of quintessence processes ($\alpha < -\frac{1}{3}$ cases) and some $F(T)$ solutions similar to ref \cite{Amir2015}, but the approach and aims of this current section are different from recent works \cite{Rodrigues2015,Amir2015,leon2,leon3}.

\subsection{Exponential ansatz solutions}\label{sect43}

By using eqn \eqref{380} exponential ansatz, we can also find FEs from eqns \eqref{401e} to \eqref{401c} as:
\begin{subequations}
\begin{align}
T &= 2c(c+2b) - \frac{2}{c_0^2}\,\exp(-2c\,t) , \label{481d}
\\
B' &= -(2c+b)+ \frac{1}{c_0^2(b-c)}\,\exp(-2c\,t), \label{481a}
\\
\kappa \rho + \frac{F(T)}{2} &= \left( T + \frac{2}{c_0^2}\,\exp(-2c\,t) \right) F_T(T), \label{481b}
\\
-\kappa \rho &=\frac{c\,F_T(T)}{\left(1+\alpha\right)} \left[ B' + c - b \right] . \label{481c}
\end{align}
\end{subequations}
In addition eqn \eqref{401a} becomes:
\begin{align}\label{481e}
\rho(t)=\frac{\rho_0}{c_0^{2\left(1+\alpha\right)}}\,\exp\left(-\left(1+\alpha\right)(b+2c)\,t\right) .
\end{align}
From eqn \eqref{481d}, we find that:
\begin{align}\label{482}
\frac{\exp(-2c\,t)}{c_0^2}= c(c + 2b)-\frac{T}{2}.
\end{align}
Then by substituting eqn \eqref{482} into eqns \eqref{481a} to \eqref{481e} and then by adding eqns \eqref{481b} and \eqref{481c}, we find that:
\begin{subequations}
\begin{align}
F(T) &=\,\frac{c\,F_T(T)}{\left(1+\alpha\right)(c-b)} \left[ T-2(c+2b)\left(b(1+2\alpha)-2\alpha\,c\right)\right] , \label{483a}
\\
\rho(T)&=\,\rho_0\,c_0^{\left(1+\alpha\right)\,\frac{b}{c}}\,\left(c(c + 2b)-\frac{T}{2}\right)^{\left(1+\alpha\right)\left(\frac{b}{2c}+1\right)} ,\label{483b}
\end{align}
\end{subequations}
where $c \neq b$, $c \neq 0$ and $\alpha \neq -1$. There are three possible cases:
\begin{enumerate}
\item ${c=-2b}$: Eqn \eqref{483a} becomes:
\begin{align}\label{484}
F(T) &=\,\frac{2}{3\left(1+\alpha\right)}\,T\,F_T(T),
\nonumber\\
& \Rightarrow\,F(T) = F(0)\,T^{\frac{3\left(1+\alpha\right)}{2}}
\end{align}
We obtain a pure power-law $F(T)$ solution as expected and eqn \eqref{483b} is:
\begin{align}\label{485}
\rho(T)&=\,\frac{\rho_0}{(-2)^{\frac{3\left(1+\alpha\right)}{4}}\,c_0^{\frac{\left(1+\alpha\right)}{2}}}\,T^{\frac{3\left(1+\alpha\right)}{4}} .
\end{align}

\item ${b=\frac{2\alpha}{(1+2\alpha)}\,c}$: Eqn \eqref{483a} becomes:
\begin{align}\label{486}
F(T) &=\,\frac{(1+2\alpha)}{\left(1+\alpha\right)} \,T\,F_T(T) ,
\nonumber\\
&\Rightarrow\,F(T)=F(0)\,T^{\frac{\left(1+\alpha\right)}{(1+2\alpha)}}.
\end{align}
where $\alpha \neq \left\lbrace-1,\, -\frac{1}{2} \right\rbrace$. Then eqn \eqref{483b} is:
\begin{align}\label{487}
\rho(T)&=\,\rho_0\,c_0^{\frac{2\alpha\left(1+\alpha\right)}{(1+2\alpha)}}\,\left(\frac{(1+6\alpha)}{(1+2\alpha)}\,c^2-\frac{T}{2}\right)^{\frac{(1+3\alpha)\left(1+\alpha\right)}{(1+2\alpha)}},
\end{align}
where $c \neq 0$ is a free parameter and $\alpha \neq \left\lbrace -1,\, -\frac{1}{2},\,-\frac{1}{3},\,0 \right\rbrace$ for a teleparallel $F(T)$ solution.

\item ${c \neq -2b}$: The general solution of eqn \eqref{483a} is:
\begin{align}
F(T) = F(0)\left[ T-2(c+2b)\left(b(1+2\alpha)-2\alpha\,c\right)\right]^{\frac{(c-b)\left(1+\alpha\right)}{c}},
\end{align}
where $b\neq c$ and $c \neq 0$ and $\rho(T)$ is eqn \eqref{483b}.

\end{enumerate}
All these previous teleparallel $F(T)$ solutions are new results. These new class of $F(T)$ solutions generalize the ref \cite{Amir2015} solutions and are simple functions.


\section{Non-linear perfect fluid solutions}\label{sect5}

After solving and finding KS solutions for a perfect isotropic linear fluid, we need to know what is happening if the perfect fluid is not linear. We will use as  matter source the perfect fluid with a non linear EoS as $P(t)=\alpha\,\rho(t)+\beta\,\left[\rho(t)\right]^w$ valid for all ${  \alpha \leq 1}$ where ${  \alpha \neq -1}$ with $w >1$ and $\beta\,\rho^{w-1} \ll \alpha$. We have in this non-linear EoS the linear dominating term plus a small power-law correction to compare with section \ref{sect4} solutions by highlighting the new terms. In the difference with ref \cite{nonvacSSpaper} and section \ref{sect4} solutions, we will find several $\alpha<-1$ teleparallel $F(T)$ solutions leading to some analytically phantom energy solutions. The eqn \eqref{302d} for conservation law becomes:  
\begin{align}
&\left[\left(1+\alpha\right)\,\rho+\beta\,\rho^w\right]\,\left(\ln(A_2\,A_3^2)\right)'+\rho'=0, 
\nonumber\\
& \Rightarrow\,\rho(t)=\frac{\rho_0}{\left[1-C\,(A_2(t)\,A_3^2(t))^{\left(1+\alpha\right)(w-1)}\right]^{\frac{1}{w-1}}},\label{501}
\end{align}
where $\rho_0=\left(-\frac{(1+\alpha)}{\beta}\right)^{\frac{1}{w-1}}$ ($\beta<0$ for a positive $\rho_0$) and $C$ is an integration constant. There are in principle an infinite number of possibilities. With the current non-linear EoS, the eqns \eqref{301a} to \eqref{301c} become:
\begin{subequations}
\begin{align}
 T =& 2\left(\ln(A_3)\right)'\left(\left(\ln(A_3)\right)'+2\left(\ln(A_2)\right)'\right) - \frac{2}{A_3^2} , \label{502a}
\\
 B' =&-\left(\ln\left(A_2\,A_3^2\right)\right)'+\frac{\frac{1}{A_3^2} - \left(\ln\left(\frac{A_2}{A_3}\right)\right)''}{\left(\ln\left(\frac{A_2}{A_3}\right)\right)' },  \label{502b}
\\
 \kappa \rho  =& \left( T + \frac{2}{A_3^2} \right) F_T(T)-\frac{F(T)}{2}, \label{502c}
\\
 -\kappa\left(1+\alpha\right)\,\rho-\kappa\beta\,\rho^w =& \left[\left(\ln(A_3)\right)'\left(B'+ \left(\ln(A_3)\right)'-\left(\ln(A_2)\right)'\right)+ \left(\ln(A_3)\right)''\right] F_T(T). \label{502d}
\end{align}
\end{subequations}
As in section \ref{sect4}, we will use similar ansatzes to solve eqns \eqref{502a} to \eqref{502d} for $A_2$, $A_3$ and $F(T)$ solutions.

\subsection{Power-law solutions}\label{sect51}

By using eqn \eqref{350} ansatz, eqn \eqref{501} becomes:
\begin{align}
\rho(t)=\frac{\rho_0}{\left[1-C_1\,t^{(b+2c)(1+\alpha)(w-1)}\right]^{\frac{1}{w-1}}} , \label{511}
\end{align}
where $b\neq -2c$ and $C_1=C\,c_0^{2\left(1+\alpha\right)(w-1)}$. For $b=-2c$, we obtain that $\rho=\rho_0=$ constant from eqn \eqref{501}. Then eqns \eqref{502a} to \eqref{502d} become:
\begin{subequations}
\begin{align}
T &= \frac{2c(c+2b)}{t^2}- \frac{2}{c_0^2}\,t^{-2c} \label{512a}
\\
B' &= \left(\frac{(1-b-2c)}{t} + \frac{t^{1-2c} }{c_0^2(b-c)} \right)  \label{512b}
\\
\kappa \rho &= \left( T + \frac{2}{c_0^2}\,t^{-2c} \right) F_T(T)-\frac{F(T)}{2}, \label{512c}
\\
-\kappa\left(1+\alpha\right)\,\rho-\kappa\beta\,\rho^w &= c\left[ \frac{B'}{t} + \frac{(c-b-1)}{t^2} \right] F_T(T). \label{512d}
\end{align}
\end{subequations}
By substituting eqns \eqref{512b} and \eqref{512c} into eqn \eqref{512d}, we find as DE:
\begin{align}
0=& \Bigg(\left( T + \frac{2}{c_0^2}\,t^{-2c} \right) F_T(T)-\frac{F(T)}{2}\Bigg)^w + \left(\frac{\kappa^{w-1}(1+\alpha)}{\beta}\right)\,\Bigg(\left( T + \frac{2}{c_0^2}\,t^{-2c} \right) F_T(T)-\frac{F(T)}{2}\Bigg)
\nonumber\\
&\,+\frac{\kappa^{w-1}\,c}{\beta}\left[ -(c+2b)\,t^{-2} + \frac{t^{-2c} }{c_0^2(b-c)}\right] F_T(T). \label{513}
\end{align}
The eqn \eqref{513} is the general and non-linear DE to solve for $F(T)$. We need to transform this eqn \eqref{513} into a solvable DE. By setting ${\bf w=2}$ in eqn \eqref{513}, we find that:
\small
\begin{align}
&\left( T + \frac{2}{c_0^2}\,t^{-2c} \right) F_T(T)-\frac{F(T)}{2}
\nonumber\\
&\quad\quad = \frac{\kappa(1+\alpha)}{2\beta}\left[-1 +\delta_1\sqrt{1-\frac{2\beta}{\kappa(1+\alpha)^2}\left[ -2c(c+2b)\,t^{-2} + \frac{2c\,t^{-2c} }{c_0^2(b-c)}\right] F_T(T)}\right]. \label{514}
\end{align}
\normalsize
where $\delta_1=\pm 1$. From eqn \eqref{512a}, we isolate the relation:
\begin{align}\label{514a}
2c(c+2b)\,t^{-2}=T +\frac{2}{c_0^2}\,t^{-2c} .
\end{align}
Then eqn \eqref{514} becomes:
\small
\begin{align}
& 2\left( T + \frac{2}{c_0^2}\,t^{-2c} \right) F_T(T)-F(T)
\nonumber\\
&\quad\quad= \frac{\kappa(1+\alpha)}{\beta}\left[-1 +\delta_1\sqrt{1+\frac{2\beta}{\kappa(1+\alpha)^2}\left[ \left(T +\frac{2}{c_0^2}\,t^{-2c}\right) - \frac{2c\,t^{-2c} }{c_0^2(b-c)}\right] F_T(T)}\right]. \label{515}
\end{align}
\normalsize
Eqn \eqref{515} is a non-linear DE and we need to simplify this equation. For a small quadratic correction (i.e, $\beta \ll (1+\alpha)$) to the linear perfect fluid EoS, eqn \eqref{515} will simplify as:
\small
\begin{align}
& 2\left( T + \frac{2}{c_0^2}\,t^{-2c} \right) F_T(T)-F(T)
\nonumber\\
&\quad\quad\approx  \frac{\kappa(1+\alpha)}{\beta}\left[\left(\delta_1-1\right) +\frac{\delta_1\,\beta}{\kappa(1+\alpha)^2}\left[ \left(T +\frac{2}{c_0^2}\,t^{-2c}\right) - \frac{2c\,t^{-2c} }{c_0^2(b-c)}\right] F_T(T)\right]. \label{516}
\end{align}
\normalsize
For $\delta_1=+1$, we obtain exactly eqn \eqref{421} for a linear perfect fluid. However for $\delta_1=-1$, eqn \eqref{516} becomes {  for $\alpha \neq -\frac{3}{2}$}:
\small
\begin{align}
F(T) - &\frac{2\kappa(1+\alpha)}{\beta}\approx  \frac{F_T(T)}{(1+\alpha)}\left[(3+2\alpha) \left(T +\frac{2}{c_0^2}\,t^{-2c}(T)\right) - \frac{2c\,t^{-2c}(T) }{c_0^2(b-c)}\right] ,
\nonumber\\ 
\Rightarrow\,F(T)\approx &\frac{2\kappa(1+\alpha)}{\beta}+\left(F(0)-\frac{2\kappa(1+\alpha)}{\beta}\right)
\nonumber\\
&\;\times\,\exp\left[(1+\alpha)\int\,dT\,\left[(3+2\alpha)\,T +\frac{2}{c_0^2}\left(3+2\alpha-\frac{c}{(b-c)}\right)\,t^{-2c}(T) \right]^{-1}\right] , \label{517}
\end{align}
\normalsize
where $\beta \neq 0$, $b \neq c$ and $c\neq 0$. We obtained at eqn \eqref{517} the general solution for possible $F(T)$. In addition, we need to solve the characteristic eqn \eqref{422} as for perfect fluids cases in section \ref{sect4} for each subcases. 

\noindent For ${\alpha=-\frac{3}{2}}$ cases, eqn \eqref{517} simplifies as:
\begin{align}
F(T)\approx &-\frac{\kappa}{\beta}+\left(F(0)+\frac{\kappa}{\beta}\right)\,\exp\left[\frac{c_0^2(b-c)}{4c}\,\int\,dT\,t^{2c}(T)\right] , \label{517s}
\end{align}
where $b \neq c$ and $c\neq 0$. The eqn \eqref{517s} is very similar to eqn \eqref{421s}, because the same integrals to solve. The only changes from $F(T)$ solutions of section \ref{sect41} are for the shifting constants depending on $\beta$.

\noindent For ${\bf c=-2b}$ case, eqn \eqref{422} simplifies as $t^{4b}(T)=\frac{c_0^2(-T)}{2}$ and then eqn \eqref{517} becomes:
\small
\begin{align}
F(T)\approx\frac{2\kappa(1+\alpha)}{\beta}+\left(F(0)-\frac{2\kappa(1+\alpha)}{\beta}\right)\,T^{-\frac{3(1+\alpha)}{2}}, \label{518}
\end{align}
\normalsize
where $\beta \neq 0$.

\noindent As for the linear perfect fluid case, we obtain the same cases ${\bf   c= \frac{1}{2},\,-\frac{1}{2},\,1,\,-1,\frac{3}{2},\,2,\,-2,\,3,}$ ${\bf  \,-3,\,4}$ for $c\neq -2b$. In this section, we will develop only the ${\bf c=\frac{1}{2},\,1,\,-1}$ and ${\bf 2}$ subcases as done in section \ref{sect4}, because all other subcases will not lead to analytic and closed $F(T)$ solution. The analytic and closed teleparallel $F(T)$ solutions are:
\begin{enumerate}
\item ${\bf c=\frac{1}{2}}$: By substituting the eqn \eqref{431} for $t^{-1}(T)$, eqn \eqref{517} becomes {  (i.e. $\alpha \neq -\frac{3}{2}$)}:
\small
\begin{align}
F(T)\approx & \frac{2\kappa(1+\alpha)}{\beta}+\left(F(0)-\frac{2\kappa(1+\alpha)}{\beta}\right)
\nonumber\\
& \times\,\Bigg[-c_0^4\left(\frac{1}{2}+2b\right)\Bigg((3+2\alpha)\,T+\frac{4}{c_0^4}\left(3+2\alpha+\frac{1}{(1-2b)}\right)
\nonumber\\
&\times\,\frac{\left[1+\delta_2\,\sqrt{1+\left(\frac{1}{2}+2b\right)\,c_0^4\,T}\right]}{\left(1+4b\right)}\Bigg)\Bigg]^{\frac{1+\alpha}{3+2\alpha}}\,
\exp\Bigg[2(1+\alpha)\left(1-2b+\frac{1}{3+2\alpha}\right)\,
\nonumber\\
&\times\,\tanh^{-1}\left(1+(1-2b)(3+2\alpha)\left(1+\delta_2\,\sqrt{1+\left(\frac{1}{2}+2b\right)\,c_0^4\,T}\right)\right)\Bigg], \label{520}
\end{align}
\normalsize
where $b \neq -\frac{1}{4}$, $\alpha \neq -\frac{3}{2}$ and $\delta_2=\pm 1$. For $\alpha=-\frac{3}{2}$, eqn \eqref{517s} becomes with eqn \eqref{431}:
\small
\begin{align}
F(T)=& -\frac{\kappa}{\beta}+\left(F(0)+\frac{\kappa}{\beta}\right)\,\left[1+\delta_1\,\sqrt{1+\left(\frac{1}{2}+2b\right)\,c_0^4\,T}\right]^{\frac{1}{2}-b}
\nonumber\\
&\,\times\,\exp\Bigg[\delta_1\,\left(b-\frac{1}{2}\right)\,\sqrt{1+\left(\frac{1}{2}+2b\right)\,c_0^4\,T}\Bigg]   ,\label{520s}
\end{align}
\normalsize
where $b \neq \frac{1}{2}$ and $\beta\neq 0$.

\item ${\bf c=1}$: By substituting the eqn \eqref{435} for $t^{-2}(T)$, eqn \eqref{517} becomes {  (i.e. $\alpha \neq -\frac{3}{2}$)}:
\begin{align}
F(T)\approx &\frac{2\kappa(1+\alpha)}{\beta}+\left(F(0)-\frac{2\kappa(1+\alpha)}{\beta}\right)\,T^{\left[ \frac{(1+\alpha)(b-1)\left(c_0^2\left(1+2b\right)-1\right)}{c_0^2(3+2\alpha)(b-1)\left(2b+1\right)-1} \right]} , \label{519}
\end{align}
where $b \neq 1$. For $\alpha=-\frac{3}{2}$, eqn \eqref{517s} becomes with eqn \eqref{435}:
\begin{align}
F(T)=-\frac{\kappa}{\beta}+\left(F(0)+\frac{\kappa}{\beta}\right)\,T^{\frac{(b-1)}{2}\,\left[(1+2b)\,c_0^2-1\right]}   , \label{519s}
\end{align}
where $b \neq 1$ and $\beta\neq 0$.

\item ${\bf c=-1}$: By substituting the eqn \eqref{437} for $t^2(T)$, eqn \eqref{517} becomes {  (i.e. $\alpha \neq -\frac{3}{2}$)}:
\small
\begin{align}
F(T)
\approx & \frac{2\kappa(1+\alpha)}{\beta}+\left(F(0)-\frac{2\kappa(1+\alpha)}{\beta}\right)
\nonumber\\
&\times\,\left[-\frac{4(3+2\alpha)}{b+1}\,T^2-\frac{16(1-2b)}{c_0^2}\left(3+2\alpha+\frac{1}{b+1}\right)^2\right]^{\frac{1}{4}\left(\frac{1+\alpha}{3+2\alpha}-(b+1)(1+\alpha)\right)}
\nonumber\\
&\times\,\left[T+\sqrt{T^2+16(1-2b)\,c_0^{-2}}\right]^{\frac{\delta_1}{2}\left(\frac{1+\alpha}{3+2\alpha}+(b+1)(1+\alpha)\right)}
\nonumber\\
&\times\,\exp\Bigg[\frac{1}{2}\left(\frac{1+\alpha}{3+2\alpha}-(b+1)(1+\alpha)\right)
\nonumber\\
&\times\,\tanh^{-1}\left[\frac{\delta_1\left(3+2\alpha-\frac{1}{b+1}\right)\,T}{\left(3+2\alpha+\frac{1}{b+1}\right)\,\sqrt{T^2+16(1-2b)\,c_0^{-2}}}\right]\Bigg] ,\label{521}
\end{align}
\normalsize
where $b \neq -1$, $\alpha \neq -\frac{3}{2}$ and $\delta_1=\pm 1$. {  For $\alpha=-\frac{3}{2}$, eqn \eqref{517s} becomes with eqn \eqref{437}:
\small
\begin{align}
F(T)=& -\frac{\kappa}{\beta}+\left(F(0)+\frac{\kappa}{\beta}\right)\,\left[T+\sqrt{T^2+16(1-2b)\,c_0^{-2}}\right]^{-\frac{\delta_1}{2}(b+1)}
\nonumber\\
&\,\times\,\exp\Bigg[-\frac{c_0^2\,(b+1)}{32(1-2b)}\,T\left(T+\delta_1\,\sqrt{T^2+16(1-2b)\,c_0^{-2}}\right)\Bigg]   ,\label{521s}
\end{align}
\normalsize
where $b \neq \left\lbrace -1,\,\frac{1}{2}\right\rbrace$ and $\beta\neq 0$. }

\item ${\bf c=2}$: By substituting the eqn \eqref{441} for $t^{-2}(T)$, eqn \eqref{517} becomes {  (i.e. $\alpha \neq -\frac{3}{2}$)}:
\small
\begin{align}
F(T)\approx &\, \frac{2\kappa(1+\alpha)}{\beta}+\left(F(0)-\frac{2\kappa(1+\alpha)}{\beta}\right)
\nonumber\\
&\times\,\left[\frac{T}{4c_0^2(1+b)^2(b-2)} +2\,\left(3+2\alpha-\frac{2}{b-2}\right)
\left[1+\delta_1\,\sqrt{1-\frac{T}{8c_0^2\,(1+b)^2}}\right]\right]^{\frac{(1+\alpha)(b-2)}{2}}
\nonumber\\
&\times\,\exp\Bigg[\delta_1(1+\alpha)\left(b-2-\frac{2}{3+2\alpha}\right)
\nonumber\\
&\times\,\tanh^{-1}\left[\frac{\delta_1\left((3+2\alpha)(b-2)-2\right)-2\sqrt{1-\frac{T}{8c_0^2\,(1+b)^2}}}{(b-2)(3+2\alpha)}\right]\Bigg] , \label{522}
\end{align}
\normalsize
where $b \neq 2$, $\alpha \neq -\frac{3}{2}$ and $\delta_1=\pm 1$. For $\alpha=-\frac{3}{2}$, eqn \eqref{517s} becomes with eqn \eqref{441}:
\small
\begin{align}
F(T)=& -\frac{\kappa}{\beta}+\left(F(0)+\frac{\kappa}{\beta}\right)\,\left[4(1+b)+4\delta_1\,\sqrt{(1+b)^2-\frac{T}{8c_0^2}}\right]^{\frac{(2-b)}{2}}
\nonumber\\
&\,\times\,\exp\Bigg[\frac{(2-b)(1+b)}{2\left((1+b)+\delta_1\,\sqrt{(1+b)^2-\frac{T}{8c_0^2}}\right)}\Bigg]   ,\label{522s}
\end{align}
\normalsize
where $b \neq 2$ and $\beta\neq 0$.

\end{enumerate}

All these previous teleparallel $F(T)$ solutions are new results. Therefore, by comparing the $F(T)$ solutions described by the eqns \eqref{421} to \eqref{442s} in section \ref{sect41} with the eqns \eqref{517} to \eqref{522s}, we see a some number of similarities. Aside from a few additional and specific constants and parameters, the \eqref{517} to \eqref{522s} of this section are related to the $F(T)$ solutions of section \ref{sect41}. One of the logical explanation is that the same characteristic equations described by eq \eqref{422} are used for both classes of $F(T)$ solutions. An additional explanation is that the DEs expressed by eqns \eqref{421} and \eqref{517} respectively have very similar terms: this consideration goes more towards the similarity of the solutions. Because there are new non-linear fluid solutions, these new $F(T)$ solutions automatically generalize more the solutions in ref \cite{Amir2015}.

\subsection{$A_2=A_3^n$ ansatz solutions}\label{sect52}

By setting $A_2=A_3^n$, we find as fluid density from eqn \eqref{501}:
\begin{align}
\rho(t)=\frac{\rho_0}{\left[1-C\,(A_3(t))^{(2+n)\left(1+\alpha\right)(w-1)}\right]^{\frac{1}{w-1}}},\label{551}
\end{align}
where $n \neq -2$. For $n=-2$, eqn \eqref{551} leads to a constant fluid density for any $A_3(t)$. Then eqns \eqref{502a} to \eqref{502d} become:
\begin{subequations}
\begin{align}
T &= 2(1+2n)\left(\ln(A_3)\right)'^2 - \frac{2}{A_3^2} \label{552a}
\\
B' &= -\frac{\left(\left(\ln(A_3)\right)''+(2+n)\left(\ln(A_3)\right)'^2 + \frac{1}{(1-n)\,A_3^2} \right)}{\left(\ln(A_3)\right)'}  \label{552b}
\\
\kappa \rho  &= \left( T + \frac{2}{A_3^2} \right) F_T(T)-\frac{F(T)}{2}, \label{552c}
\\
-\kappa\left(1+\alpha\right)\,\rho-\kappa\beta\,\rho^w &= \left[ \left(\ln(A_3)\right)' B' + \left(\ln(A_3)\right)''+(1-n)\left(\ln(A_3)\right)'^2 \right] F_T(T). \label{552d}
\end{align}
\end{subequations} 
By substituting eqns \eqref{552b} and \eqref{552c} into eqn \eqref{552d}, we obtain as DE:
\begin{align}
0=& \Bigg[\left( T + \frac{2}{A_3^2} \right) F_T(T)-\frac{F(T)}{2}\Bigg]^w+\frac{\left(1+\alpha\right)\,\kappa^{w-1}}{\beta}\,\Bigg[\left( T + \frac{2}{A_3^2} \right) F_T(T)-\frac{F(T)}{2}\Bigg]
\nonumber\\
&\, - \frac{\kappa^{w-1}}{\beta}\left[(1+2n)\left(\ln(A_3)\right)'^2+\frac{1}{(1-n)\,A_3^2}\right] F_T(T). \label{553}
\end{align}
The eqn \eqref{553} is the general and non-linear DE to solve for $F(T)$. From eqn \eqref{552a}, we can isolate the expression:
\begin{align}\label{554}
(1+2n)\left(\ln(A_3)\right)'^2 = \frac{1}{2}\left( T + \frac{2}{A_3^2}\right),
\end{align}
and then eqn \eqref{553} becomes:
\begin{align}
0=& \Bigg[\left( T + \frac{2}{A_3^2} \right) F_T(T)-\frac{F(T)}{2}\Bigg]^w+\frac{\left(1+\alpha\right)\,\kappa^{w-1}}{\beta}\,\Bigg[\left( T + \frac{2}{A_3^2} \right) F_T(T)-\frac{F(T)}{2}\Bigg]
\nonumber\\
&\, - \frac{\kappa^{w-1}}{2\beta}\left[\left( T + \frac{2}{A_3^2}\right)+\frac{2}{(1-n)\,A_3^2}\right] F_T(T). \label{555}
\end{align}
For ${\bf w=2}$ fluid case and in the situation where $\beta \ll 1+\alpha$, eqn \eqref{555} will become:
\small
\begin{align}
&\left( T + \frac{2}{A_3^2} \right) F_T(T)-\frac{F(T)}{2} 
\nonumber\\
&\quad= \frac{\left(1+\alpha\right)\,\kappa}{2\beta}\left[-1+\delta_1\sqrt{1+\frac{2\beta}{\kappa\,\left(1+\alpha\right)^2}\left[\left( T + \frac{2}{A_3^2}\right)+\frac{2}{(1-n)\,A_3^2}\right] F_T(T)}\right] ,
 \nonumber\\
&\quad \approx  \frac{\left(1+\alpha\right)\,\kappa}{2\beta}\left(\delta_1-1\right)+\frac{\delta_1}{2\left(1+\alpha\right)}\left[\left( T + \frac{2}{A_3^2}\right)+\frac{2}{(1-n)\,A_3^2}\right] F_T(T).  \label{556}
\end{align}
\normalsize
For $\delta_1=+1$, we obtain exactly the eqn \eqref{404} for linear perfect fluid. We will solve only for $\delta_1=-1$ situation and eqn \eqref{556} becomes:
\small
\begin{align}
F(T) \approx &\frac{2\left(1+\alpha\right)\,\kappa}{\beta}+\frac{1}{\left(1+\alpha\right)}\left[(3+2\alpha)\,T+\frac{4}{A_3^2\,(1-n)}\left((2+\alpha)-n\left(\frac{3}{2}+\alpha\right)\right)\right] F_T(T) \label{557}
\end{align}
\normalsize
For a pure $F(T)$ solution valid for all $A_3$, we need to set $n=\frac{2+\alpha}{\frac{3}{2}+\alpha}$ where $n \neq 1$ and $\alpha \neq -\frac{3}{2}$. From eqn \eqref{557}, there are two possible situations:
\begin{enumerate}
\item ${  \alpha \neq -\frac{3}{2}}$ general case: Eqn \eqref{557} simplifies and the solution is:
\begin{align}\label{558}
F(T) \approx &\frac{2\left(1+\alpha\right)\,\kappa}{\beta}+\frac{(3+2\alpha)}{\left(1+\alpha\right)}\,T\, F_T(T) ,
\nonumber\\
\Rightarrow\,F(T) \approx & \frac{2\left(1+\alpha\right)\,\kappa}{\beta}+\left(F(0)-\frac{2\left(1+\alpha\right)\,\kappa}{\beta}\right)\,T^{\frac{\left(1+\alpha\right)}{2\,\left(\frac{3}{2}+\alpha\right)}} .
\end{align}

\item ${  \alpha = -\frac{3}{2}}$: We set $\alpha=-\frac{3}{2}+\Delta\alpha$ for studying $F(T)$ solutions around $\alpha = -\frac{3}{2}$ (where $\Delta\alpha\,\rightarrow\,0$). Then eqn \eqref{557} becomes:
\small
\begin{align}
F(T) \approx &\frac{\left(-1+2\Delta\alpha\right)\,\kappa}{\beta}-4\left[\Delta\alpha\,T+\frac{\left(1-2(n-2)\Delta\alpha\right)}{A_3^2\,(1-n)}\right] F_T(T). \label{558a}
\end{align}
\normalsize
If $n\,\rightarrow\,\infty$ and $\Delta\alpha\,\rightarrow\,0$, we obtain a GR solution. For an $A_3$ independent solution, we need to satisfy $\Delta\alpha \approx \frac{1}{2(n-2)}$ where $n \gg 1$, and then eqn \eqref{558a} becomes:
\begin{align}
F(T) \approx &-\frac{\kappa}{\beta}-\frac{2}{(n-2)}\,T\, F_T(T), 
\nonumber\\
\Rightarrow\,& F(T) \approx -\frac{\kappa}{\beta}+\left(F(0)+\frac{\kappa}{\beta}\right)\,T^{1-\frac{n}{2}}.\label{558b}
\end{align}
We find at eqn \eqref{558b} a finite limit of $F(T)$ valid for large $n$. If there is a singularity at $\alpha=-\frac{3}{2}$, the $F(T)$ solution is well defined close to this point.

\end{enumerate}

For finding $A_3$ with $\alpha \neq -\frac{3}{2}$, we need to put eqns \eqref{551} and \eqref{552c} together and then find a DE for $w=2$ fluid case. Then by substituting eqn \eqref{558} inside, we find that:
\small
\begin{align}\label{559}
0=\left(\frac{\kappa}{\beta}\right) \frac{\beta\rho_0 +\left(1+\alpha\right)\left(1-C\,A_3^{(2+n)\left(1+\alpha\right)}\right)}{\left(F(0)-\frac{2\left(1+\alpha\right)\,\kappa}{\beta}\right)\left(1-C\,A_3^{(2+n)\left(1+\alpha\right)}\right)}  - \frac{2}{A_3^2} \left(\frac{1+\alpha}{3+2\alpha}\right)\,T^{\frac{\left(1+\alpha\right)}{\left(3+2\alpha\right)}-1}+\left(\frac{1}{2}-\frac{1+\alpha}{3+2\alpha}\right)\,T^{\frac{\left(1+\alpha\right)}{\left(3+2\alpha\right)}} .
\end{align}
\normalsize
This eqn \eqref{559} is in principle hard to solve. However, there are some specific values of $\alpha$ (linear EoS parameter) where $\alpha<-1$ (phantom energy cases) leading to analytical solutions for $A_3$:
\begin{enumerate}
\item ${  \alpha=-\frac{4}{3}}$: Eqn \eqref{559} becomes:
\begin{align}\label{560}
0=&\left(\frac{\kappa\,A_3^2}{2\beta}\right) \frac{\beta\rho_0 -\frac{1}{3}\left(1-C\,A_3^{-2}\right)}{\left(F(0)+\frac{2\kappa}{3\beta}\right)\left(1-C\,A_3^{-2}\right)} +\frac{3\,A_3^2}{4}\,T^{-1}  + T^{-2}, 
\nonumber\\
&\Rightarrow\,T=\Bigg[-\frac{3\,A_3^2}{8}+\delta_1\sqrt{\frac{9\,A_3^4}{64}-\left(\frac{\kappa\,A_3^2}{2\beta}\right) \frac{\beta\rho_0 -\frac{1}{3}\left(1-C\,A_3^{-2}\right)}{\left(F(0)+\frac{2\kappa}{3\beta}\right)\left(1-C\,A_3^{-2}\right)}}\Bigg]^{-1},
\end{align}
where $\delta_1=\pm 1$. Eqn \eqref{552a} will be in terms of $\alpha$:
\begin{align}\label{561}
T &= \left(\frac{11+6\alpha}{\frac{3}{2}+\alpha}\right)\frac{A_3'^2}{A_3^2} - \frac{2}{A_3^2}.
\end{align}
Then by putting eqns \eqref{560} and \eqref{561} together for $\alpha=-\frac{4}{3}$, we find the DE to solve for $A_3$:
\begin{align}\label{562}
3\delta_2\,A_3'=\Bigg[1-\frac{4}{3}\Bigg[1-\delta_1\sqrt{1+\frac{32}{27\,A_3^2}\left(\frac{\kappa}{\beta\,\tilde{F}(0)}\right) \left(1-\frac{3\beta\rho_0}{\left(1-C\,A_3^{-2}\right)}\right)}\Bigg]^{-1}\Bigg]^{1/2},
\end{align}
where $\tilde{F}(0)=F(0)+\frac{2\kappa}{3\beta}$ is an effective constant. This integral described by eqn \eqref{562} is very complex and there is no exact solution. Therefore there are two solvable limits leading to approximated $A_3$ solutions:
\begin{enumerate}
\item $\kappa\,\rho_0 \ll \tilde{F}(0)$ (low density limit): Eqn \eqref{562} will be approximated as:
\small
\begin{align}\label{563}
\frac{\delta_2\,(t-t_0)}{3\sqrt{3}}\approx &\int\, \left[\frac{\delta_1\sqrt{\frac{32}{27}\left(\frac{\kappa}{\beta\,\tilde{F}(0)}\right)+A_3^2}-A_3}{3\delta_1\sqrt{\frac{32}{27}\left(\frac{\kappa}{\beta\,\tilde{F}(0)}\right)+A_3^2}+A_3}\right]^{1/2}\Bigg[1+\left(\frac{\kappa}{\beta\,\tilde{F}(0)}\right)\,\frac{32\delta_1\,A_3^3}{9\left(C-A_3^2\right)}
\nonumber\\
&\;\times\,\left(\frac{32}{27}\left(\frac{\kappa}{\beta\,\tilde{F}(0)}\right)+A_3^2\right)^{-1/2}\left(\delta_1\sqrt{\frac{32}{27}\left(\frac{\kappa}{\beta\,\tilde{F}(0)}\right)+A_3^2}-A_3\right)^{-1}
\nonumber\\
&\;\times\, \left(3\delta_1\sqrt{\frac{32}{27}\left(\frac{\kappa}{\beta\,\tilde{F}(0)}\right)+A_3^2}+A_3\right)^{-1}\Bigg] \,dA_3.
\end{align}
\normalsize
This eqn \eqref{563} is very complex to solve and will not give an exact solution. But there are two limit cases for $\delta_1=-1$:
\begin{itemize}
\item $\frac{\kappa}{\beta\,\tilde{F}(0)}\ll 1$: Eqn \eqref{563} will be approximated as:
 \begin{align}\label{564}
\frac{\delta_2\,(t-t_0)}{3\sqrt{3}}\approx & A_3+\frac{8\kappa}{27\beta\,\tilde{F}(0)}\left(\frac{1}{A_3}-\frac{3}{\sqrt{C}}\tanh^{-1}\left(\frac{A_3}{\sqrt{C}}\right)\right),
\nonumber\\
\approx & A_3 + f_1(A_3),
\end{align}
where $f_1(A_3)\ll A_3$. In the $\frac{\kappa}{\beta\,\tilde{F}(0)}\,\rightarrow\,0$ limit, we find that $A_3=\frac{\delta_2\,(t-t_0)}{3\sqrt{3}}$ as for vacuum solution. The function $A_3$ is bounded by $0 < A_3 < +\sqrt{C}$ (or $-\sqrt{C}< A_3< 0$) and the limits on $f_1(A_3)$ are :
\begin{itemize}
\item $A_3\,\rightarrow 0^{+}$  and $A_3\,\rightarrow -\sqrt{C}$: $f_1(A_3)\,\rightarrow\,+\infty$.

\item $A_3\,\rightarrow +\sqrt{C}$ and $A_3\,\rightarrow 0^{-}$: $f_1(A_3)\,\rightarrow\,-\infty$.
\end{itemize} 
With a $f_1(A_3)$ can be going to infinity, this case of solution leads to an unstable universe.

\item $\frac{\kappa}{\beta\,\tilde{F}(0)}\gg 1$: Eqn \eqref{563} will be simplified as:
\begin{align}\label{565}
\frac{\delta_2\,(t-t_0)}{3}\approx & A_3+ \frac{1}{8}\,\sqrt{\frac{3\beta\,\tilde{F}(0)}{2\kappa}}\left(3C\,\ln\left(A_3^2-C\right)+5\,A_3^2\right),
\nonumber\\
\approx & A_3+f_2(A_3),
\end{align}
where $f_2(A_3)\ll A_3$. For the $\frac{\kappa}{\beta\,\tilde{F}(0)}\,\rightarrow\,0$ limit, we still find that $A_3=\frac{\delta_2\,(t-t_0)}{3\sqrt{3}}$ as previously. The $f_2(A_3)$ correction function is defined for $A_3 <-\sqrt{C}$ and $A_3>+\sqrt{6}$ only and the limits are:
\begin{itemize}
\item $A_3\,\rightarrow\,\pm \infty$: $f_2(A_3)\,\rightarrow\,+\infty$.

\item $A_3\,\rightarrow\,\pm \sqrt{C}$: $f_2(A_3)\,\rightarrow\,-\infty$.
\end{itemize}
With a $f_2(A_3)$ can be going to infinity, this case of solution leads to an unstable universe.

\end{itemize}

\item $\kappa\,\rho_0 \gg \tilde{F}(0)$ (high density limit):  Eqn \eqref{562} will be approximated as:
\begin{align}\label{566}
\frac{\delta_2\,(t-t_0)}{3}\approx &\,A_3-\frac{\delta_1}{4}\left(\frac{\tilde{F}(0)}{2\kappa\,\rho_0}\right)^{1/2}\,\left(A_3\,\sqrt{C-A_3^2}+C\,\arctan\left(\frac{A_3}{\sqrt{C-A_3^2}}\right)\right),
\nonumber\\
\approx & A_3+ g(A_3),
\end{align}
where $g(A_3)\ll A_3$. The eqn \eqref{566} is a bounded equation because there is an inverse trigonometric function ($-\sqrt{C} < A_3 < +\sqrt{C}$). In the $\frac{\tilde{F}(0)}{2\kappa\,\rho_0}\,\rightarrow\,0$ limit, we find that $A_3\,\rightarrow\,\frac{\delta_2\,(t-t_0)}{3}$ as for vacuum solution. The $g(A_3)$ correction to $A_3$ will be for the both limits of $A_3$:
\begin{itemize}
\item $A_3\,\rightarrow\,+\sqrt{C}$: $g(A_3)=-\frac{\delta_1\,C\,\pi}{8}\left(\frac{\tilde{F}(0)}{2\kappa\,\rho_0}\right)^{1/2}$.

\item $A_3\,\rightarrow\,-\sqrt{C}$: $g(A_3)=+\frac{\delta_1\,C\,\pi}{8}\left(\frac{\tilde{F}(0)}{2\kappa\,\rho_0}\right)^{1/2}$.
\end{itemize}
The $A_3$ solution is bounded by two linear functions of $t$ as:
\begin{align}
\frac{\delta_2\,(t-t_0)}{3}-\frac{\delta_1\,C\,\pi}{8}\left(\frac{\tilde{F}(0)}{2\kappa\,\rho_0}\right)^{1/2} < A_3(t) < \frac{\delta_2\,(t-t_0)}{3}+\frac{\delta_1\,C\,\pi}{8}\left(\frac{\tilde{F}(0)}{2\kappa\,\rho_0}\right)^{1/2}.
\end{align}

\end{enumerate}

\item ${  \alpha=-\frac{5}{3}}$: Eqn \eqref{559} becomes in this case:
\begin{align}\label{570}
0=&-\left(\frac{2\kappa}{3\beta}\right) \frac{\beta\rho_0 -\frac{2}{3}\left(1-C\right)}{\left(F(0)+\frac{4\,\kappa}{3\beta}\right)\left(1-C\right)}  + \frac{8}{3\,A_3^2}\,T+T^{2} ,
\nonumber\\
&\Rightarrow\,T= -\frac{4}{3\,A_3^2}+\delta_1 \sqrt{\frac{16}{9\,A_3^4}+\left(\frac{2\kappa}{3\beta}\right) \frac{\beta\rho_0 -\frac{2}{3}\left(1-C\right)}{\left(F(0)+\frac{4\,\kappa}{3\beta}\right)\left(1-C\right)} }  ,
\end{align}
where $\delta_1=\pm 1$. By using eqn \eqref{561} for $\alpha=-\frac{5}{3}$ case and then by merging to eqn \eqref{570}, we find the DE to solve for $A_3$:
\begin{align}\label{571}
3\delta_2\,A_3' \approx \Bigg[-1-2\delta_1\,\sqrt{1-\left(\frac{\kappa}{8\beta\,\tilde{F}_0}\right)\left(2+\frac{3\beta\rho_0}{\left(C-1\right)}\right)\,A_3^4} \Bigg]^{1/2} ,
\end{align}
where $\tilde{F}(0)=F(0)+\frac{4\kappa}{3\beta}$ here. The eqn \eqref{571} is a very complex integral without exact solution. But there are two solvable limits leading to approximated $A_3$ solutions:
\begin{enumerate}
\item $\kappa\,\rho_0 \ll \tilde{F}(0)$ (low density limit): Eqn \eqref{571} will be approximated as:
\begin{align}\label{572}
\frac{\delta_2\,(t-t_0)}{3} \approx & \int\,\Bigg[\frac{1}{\sqrt{-2\delta_1\sqrt{1-\frac{\kappa}{4\beta\,\tilde{F}(0)}\,A_3^4}-1}}+\frac{3\delta_1\kappa\,\rho_0}{16\tilde{F}(0)^2(1-C)}
\nonumber\\
&\;\times\,\frac{A_3^4}{\sqrt{1-\frac{\kappa}{4\beta\,\tilde{F}(0)}\,A_3^4}\left(-2\delta_1\sqrt{1-\frac{\kappa}{4\beta\,\tilde{F}(0)}\,A_3^4}-1\right)^{3/2}}\Bigg]\,dA_3.
\end{align}
This eqn \eqref{572} is complex to solve and needs some specific approximations. There are two possible subcases:
\begin{itemize}
\item $\frac{\kappa}{\beta\,\tilde{F}(0)}\ll 1$: Eqn \eqref{572} becomes for $\delta_1=-1$ (real number solution):
\small
\begin{align}\label{573}
 & \frac{\delta_2\,(t-t_0)}{3} \approx  A_3+ \frac{\kappa}{40\,\beta\,\tilde{F}(0)}\left(1+\frac{3\beta\,\rho_0}{2\tilde{F}(0)(C-1)}\right)\,A_3^5,
\nonumber\\
\Rightarrow &\,A_3(t)=\frac{\delta_2\,(t-t_0)}{3}
\nonumber\\
&\times\,_4F_3\left(\frac{1}{5},\frac{2}{5},\frac{3}{5},\frac{4}{5};\frac{1}{2},\frac{3}{4},\frac{5}{4};-\left(\frac{5}{12}\right)^4\frac{\kappa}{8\beta\,\tilde{F}(0)}\left(1+\frac{3\beta\,\rho_0}{2\tilde{F}(0)(C-1)}\right)(t-t_0)^4\right).
\end{align}
\normalsize

\item $\frac{\kappa}{\beta\,\tilde{F}(0)}\gg 1$: Eqn \eqref{572} becomes for $\delta_1=-1$:
\begin{align}\label{574}
&\frac{\delta_2\,(t-t_0)}{3} \approx  \left(1+\frac{3\beta\,\rho_0}{8\tilde{F}(0)(1-C)}\right)\left(-\frac{\beta\,\tilde{F}(0)}{\kappa}\right)^{1/4}\,\ln\,A_3,
\nonumber\\
&\Rightarrow\,A_3(t)=A_3(0)\,\exp\left[\left(-\frac{\kappa}{\beta\,\tilde{F}(0)}\right)^{1/4}\,\frac{\delta_2}{3\left(1+\frac{3\beta\,\rho_0}{8\tilde{F}(0)(1-C)}\right)}\,t\right].
\end{align}
We find on eqn \eqref{574} an exponential function for $A_3$. We must have $\tilde{F}(0) < 0$ for a real $A_3$ and the universe is:
\begin{itemize}
\item Expanding: $\delta_2=+1$ and $C<1$ or $\delta_2=-1$ and $C>1$.

\item Contracting: $\delta_2=-1$ and $C<1$ or $\delta_2=+1$ and $C>1$.
\end{itemize}

\end{itemize}

\item $\kappa\,\rho_0 \gg \tilde{F}(0)$ (high density limit): Eqn \eqref{571} will be approximated as:
\begin{align}\label{576}
\frac{\delta_2\,(t-t_0)}{3} \approx & \sqrt{-\delta_1}\,\left(\frac{2\tilde{F}(0)^2\,(1-C)}{3\kappa\,\rho_0}\right)^{1/4}\,\ln\,A_3.
\end{align}
For $\delta_1=1$, eqn \eqref{576} leads to an oscillating $A_3$ solution, which is not physically relevant. For $\delta_1=-1$, we find an exponential $A_3$ solution:
\begin{align}\label{577}
& \frac{\delta_2\,(t-t_0)}{3} \approx \sqrt{-\delta_1}\,\left(\frac{2\tilde{F}(0)^2\,(1-C)}{3\kappa\,\rho_0}\right)^{1/4}\,\ln\,A_3,
\nonumber\\
&\Rightarrow\,A_3(t)=A_3(0)\,\exp\left[\frac{\delta_2}{3}\left(\frac{3\kappa\,\rho_0}{2\tilde{F}(0)^2\,(1-C)}\right)\,t\right].
\end{align}
We have the following scenarios for universe evolution:
\begin{itemize}
\item Expanding: $\delta_2=+1$ and $C<1$ or $\delta_2=-1$ and $C>1$.

\item Contracting: $\delta_2=-1$ and $C<1$ or $\delta_2=+1$ and $C>1$.
\end{itemize}

\end{enumerate}

\end{enumerate}

All these previous teleparallel $F(T)$ solutions are really new results. However, the solutions applicable to phantom energy and Big Rip models are not covered in ref \cite{Amir2015}. Only on this point, the $F(T)$ solutions newly obtained in this section constitute a very nice advance specific to non-linear perfect fluids with a positive impact on the teleparallel $F(T)$ gravity possibilities. These new results highlight the differences between the solutions for linear EoS and those for non-linear EoS in the teleparallel gravity context. The main difference lies in the types of physical processes targeted: non-linear perfect fluids lead more easily to the study of phantom energy and Big Rip processes (where $\alpha < -1$), which linear perfect fluids will not do this in general (because usually limited to $-1< \alpha < 1$). This difference can be seen by comparing the possible subcases of $F(T)$ solutions between sections \ref{sect42} and \ref{sect52}. Indeed, the quadratic correction term $\beta \rho^2$ in the non-linear EoS induces this difference.

\subsection{Exponential ansatz solutions}\label{sect53}

Still by using eqn \eqref{380} exponential ansatz, we can find FES from eqns \eqref{502a} to \eqref{502d} as:
\begin{subequations}
\begin{align}
T &= 2c(c+2b) - \frac{2}{c_0^2}\,\exp(-2c\,t) , \label{581a}
\\
B' &= -(2c+b)+ \frac{1}{c_0^2(b-c)}\,\exp(-2c\,t), \label{581b}
\\
\kappa \rho  &= \left( T + \frac{2}{c_0^2}\,\exp(-2c\,t) \right) F_T(T)-\frac{F(T)}{2}, \label{581c}
\\
-\kappa\left(1+\alpha\right)\,\rho-\kappa\beta\,\rho^w &= c\,\left[B' + c - b \right] F_T(T). \label{581d}
\end{align}
\end{subequations}
In addition, eqn \eqref{501} becomes:
\begin{align}\label{581e}
\rho(t)=\rho_0\,\left[1-C\,( \,c_0^2)^{\left(1+\alpha\right)(w-1)}\,\exp\left(\left(1+\alpha\right)(w-1)(b+2c)\,t\right)\right]^{-\frac{1}{w-1}}.
\end{align}
From eqn \eqref{581a} we find as solution eqn \eqref{482}, then by substuting eqns \eqref{482} and \eqref{581b} into eqns \eqref{581c} to \eqref{581e} and by substituting eqn \eqref{581c} into \eqref{581d}, we obtain that:
\begin{subequations}
\begin{align}
&-\left(1+\alpha\right)\,\Bigg(2c(c+2b) F_T(T)-\frac{F(T)}{2}\Bigg)-\kappa^{1-w}\beta\,\Bigg(2c(c+2b) F_T(T)-\frac{F(T)}{2}\Bigg)^w
\nonumber\\
 &\quad\quad = \frac{c}{(b-c)}\,\left[(c+2b)(2c-b)-\frac{T}{2}\right] F_T(T), \label{582a}
\\
\rho(T)&=\rho_0\,\left[1-C\,c_0^{-\left(1+\alpha\right)(w-1)\frac{b}{c}}\left( c(c + 2b)-\frac{T}{2}\right)^{-\left(1+\alpha\right)(w-1)\left(\frac{b}{2c}+1\right)}\right]^{-\frac{1}{w-1}}.\label{582b}
\end{align}
\end{subequations}
There are two possible cases of solutions for eqn \eqref{582a}:
\begin{enumerate}
\item ${\bf c=-2b}$: Eqn \eqref{582a} simplifies as a simple DE:
\begin{align}
& F(T)\left[1+\frac{\beta}{(-2\kappa)^{w-1}\left(1+\alpha\right)}\,\left(F(T)\right)^{w-1}\right] = \frac{2}{3\left(1+\alpha\right)}\,T\,F_T(T), \label{583}
\end{align}
By integration, the solution of eqn \eqref{583} is:
\begin{align}\label{584}
F(T)=\Bigg[\left(\frac{\beta}{(-2\kappa)^{w-1}\left(1+\alpha\right)}+\left(F(0)\right)^{1-w}\right)\,T^{-\frac{3(w-1)\left(1+\alpha\right)}{2}}-\frac{\beta}{(-2\kappa)^{w-1}\left(1+\alpha\right)}\Bigg]^{\frac{1}{1-w}},
\end{align}
where $w>1$, $b\neq c$ and $c \neq 0$.

\item ${\bf c \neq -2b}$ (general case): We can solve eqn \eqref{582a} for $w=2,\,3,$ and $4$. However, the $w=3$ and $4$ cases are complex to solve and we will restrict to $w=2$ subcase. Then eqn \eqref{582a} becomes in this case:
\begin{align}\label{585}
0=&\Bigg(2c(c+2b) F_T(T)-\frac{F(T)}{2}\Bigg)^2+\frac{\kappa\left(1+\alpha\right)}{\beta}\,\Bigg(2c(c+2b) F_T(T)-\frac{F(T)}{2}\Bigg)
\nonumber\\
&+ \frac{c\,\kappa}{(b-c)\,\beta}\,\left[(c+2b)(2c-b)-\frac{T}{2}\right] F_T(T), 
\nonumber\\
\Rightarrow\,&  4c(c+2b) F_T(T)-F(T)
\nonumber\\
&\quad\quad\,= \frac{\kappa\left(1+\alpha\right)}{\beta}\left[-1+\delta_1\sqrt{1+\frac{2\beta c}{(c-b)\kappa\left(1+\alpha\right)^2}\,\left(2(c+2b)(2c-b)-T\right) F_T(T)}\right].
\end{align}
Eqn \eqref{585} is a non-linear DE and we need to approximate this relation for $\beta \ll (1+\alpha)$, the weak quadratic term approximation. In this case, eqn \eqref{585} will be approximated as:
\small
\begin{align}\label{586}
&4c(c+2b) F_T(T)-F(T) \approx \frac{\kappa\left(1+\alpha\right)}{\beta}(\delta_1-1)+\frac{\delta_1\,c}{(c-b)\left(1+\alpha\right)}\,\left(2(c+2b)(2c-b)-T\right) F_T(T) .
\end{align}
\normalsize
For $\delta_1=+1$, eqn \eqref{586} leads exactly to eqn \eqref{483a} for linear perfect fluids with eqn \eqref{483b} as $\rho(T)$. For $\delta_1=-1$, eqn \eqref{586} becomes the DE:
\begin{align}\label{587}
& F(T)-\frac{2\kappa\left(1+\alpha\right)}{\beta}\approx \frac{c\,F_T(T)}{(b-c)\left(1+\alpha\right)}\,\left[T-2(c+2b)\left(2(\alpha+2)\,c-(3+2\alpha)\,b\right)\right] .
\end{align}
There are two possible subcases for eqn \eqref{587} solutions when $c \neq -2b$:
\begin{enumerate}
\item ${  b=\frac{2(2+\alpha)}{(3+2\alpha)}\,c}$ case: Eqn \eqref{587} simplifies:
\begin{align}\label{588}
& F(T)-\frac{2\kappa\left(1+\alpha\right)}{\beta} \approx\frac{(3+2\alpha)}{\left(1+\alpha\right)}\,T\,F_T(T),
\nonumber\\
& \Rightarrow\,F(T) \approx \frac{2\kappa\left(1+\alpha\right)}{\beta}+\left(F(0)-\frac{2\kappa\left(1+\alpha\right)}{\beta}\right)\,T^{\frac{\left(1+\alpha\right)}{(3+2\alpha)}} .
\end{align}
Then eqn \eqref{582b} for the fluid density is:
\begin{align}
\rho(T)&=\rho_0\,\left[1-C\,c_0^{-\frac{2\left(1+\alpha\right)(2+\alpha)}{(3+2\alpha)}}\left( c^2\,\frac{(11+6\alpha)}{(3+2\alpha)}-\frac{T}{2}\right)^{-\frac{\left(1+\alpha\right)(5+3\alpha)}{(3+2\alpha)}}\right]^{-1}.\label{589}
\end{align}

\item General case (${c \neq -2b}$ and ${b \neq \frac{2(2+\alpha)}{(3+2\alpha)}\,c}$): By integrating eqn \eqref{587}, we obtain that:
\begin{align}\label{590}
F(T) \approx \frac{2\kappa\left(1+\alpha\right)}{\beta}+\left[F(0)-\frac{2\kappa\left(1+\alpha\right)}{\beta}\right]\,\left[T-T_0\right]^{\frac{(b-c)\left(1+\alpha\right)}{c}},
\end{align}
where $T_0=2(c+2b)\left(2(\alpha+2)\,c-(3+2\alpha)\,b\right)$ is a constant and $F(0)$ is the integration constant. The fluid density $\rho(T)$ is described by eqn \eqref{582b}.

\end{enumerate}

\end{enumerate}
All these previous teleparallel $F(T)$ solutions are new results. Once again, these new class of $F(T)$ solutions generalizes the ref \cite{Amir2015} solutions because these are non-linear EoS $F(T)$ solutions. In addition, the $F(T)$ solutions obtained in the present section are almost identical to those in section \ref{sect43}, apart from a shift taking into account the quadratic correction in the EoS. Once again, the solutions freshly obtained here are similar to those obtained in section \ref{sect43}.


\section{Discussion and Conclusion}\label{sect6}

The main aim of this paper was to obtain KS teleparallel $F(T)$ solutions for vacuum, linear and non-linear perfect fluids. This was achieved  through the use of various ansatzes, for example power-law and exponential (infinite superposition of power-laws). We obtained not only power-law $F(T)$ solutions, but also more complex $F(T)$ as we can see in sections \ref{sect41} and \ref{sect51} for $ c=\frac{1}{2},\,1,\,-1,\,2$ via power-law ansatz solutions. As mentioned before, these teleparallel $F(T)$ solutions found in sections \ref{sect41} and \ref{sect51} by this ansatz have many similar terms and were also obtained by the same form of characteristic equation described by eqn \eqref{422}. Obviously for $c=1$, we find a power-law $F(T)$ solution generalizing the vacuum solution. For vacuum case, we found that $A_3$ component can only be linear in $t$, which limits the possible analytical solutions. For the exponential ansatz, the teleparallel $F(T)$ solutions appear only for perfect fluids in sections \ref{sect43} and \ref{sect53} and they are power-law solutions (no additional constant term for linear and an additional constant term for non-linear): this situation represents an infinite sum limit of power-law terms.

However, some special solutions are found in sections \ref{sect42} and \ref{sect52}. We find power-law $F(T)$ solutions as before, therefore there are various types of solutions appearing for $A_3$. For linear perfect fluids $P=\alpha\,\rho$ in section \ref{sect42}: we obtain an exact solution for $A_3$ when $\alpha=-\frac{1}{3}$, but we must approximate the solutions for low and high cosmological fluid densities situations for the cases $\alpha=-\frac{1}{4}$ and $-\frac{2}{3}$. For low densities, these are relevant models for predicting the future evolution of an expanding universe. For high densities, these models are especially useful for explaining the universe just after the big bang. In some specific cases, we find as the limit a linear $A_3$ in $t$ as for the vacuum situation. All this without taking into account that for $\alpha< -\frac{1}{3}$ cases, we are able to model the famous quintessence process. In the latter case, it will be necessary to replace the perfect fluid by a scalar field to achieve this \cite{steinhardt1,steinhardt2,caldwell1}. The study of this physical process with the teleparallel $F(T)$ solutions found in this paper would be necessary for a more complete theory on quintessence.

In the case of a non-linear perfect fluids in section \ref{sect52}, we obtain power-law $F(T)$ solutions with an additional constant term. However, we extend the perfect fluid linear term $\alpha$ definition to $\alpha<-1$ values for finding the analytically solvable $A_3$ solutions. We find that solutions for $\alpha=-\frac{4}{3}$ and $\alpha=-\frac{5}{3}$ describe the phantom energy cases (negative kinetic energy) which could lead at the end of evolution to the Big Rip according to some recent works \cite{farnes,baumframpton}. In these latter cases, we carried out the high and low fluid density approximations for the $A_3$ solutions for the possible limits. But, we also studied the stability of correction terms with respect to the dominant term, a bit as in refs. \cite{attractor,Kofinas}. We notice that for the case $\alpha=-\frac{4}{3}$ at high fluid density, we obtain a $A_3$ with linear functions as minimal and maximal limits: which means a stable solution for $A_3 $. All other subcases do not offer finite limits for $A_3$, so these solutions may be divergent and the universe model is therefore unstable. Once again, a more detailed study is really necessary on this phenomenon of phantom energy and the Big Rip process in teleparallel $F(T)$ gravity. These two phenomenons are some critical and concern dark energy in the universe. They deserve better answers. Nevertheless, the teleparallel $F(T)$ gravity approach has proved powerful enough to provide this new class of solutions for achieving them in the future.

Apart from the previous cases concerning universe models, there were two recurring situations in section \ref{sect42} and \ref{sect52} of non-possibility for $F(T)$ solution when $\alpha=-\frac{1}{2}$ for linear EoS and $\alpha=-\frac{3}{2}$ for quadratic EoS respectively. The problem does not arise by using power-law ansatz solutions in sections \ref{sect41} and \ref{sect51} for $\alpha=-\frac{1}{2}$ and $-\frac{3}{2}$ subcases respectively: there are several $F(T)$ solutions as seen previously. Therefore, in section \ref{sect42} and \ref{sect52} situations, we have proved by approximations on $\alpha$ close to $-\frac{1}{2}$ and $-\frac{3}{2}$ respectively that $F(T)$ described by eqns \eqref{406} and \eqref{558b} are defined all around these both critical values. There are only ansatz caused singularities because the value of $n$ in this ansatz goes to infinity at $\alpha=-\frac{1}{2}$ and $-\frac{3}{2}$. There are only a purely mathematical concern, not physical in both situations.

After the considerations on dark energy, quintessence and Big Rip models, there are some possible works for finding KS teleparallel $F(T)$ solutions for electromagnetic sources. This possible future works can be useful for studying spacetimes with electrically charged particle sources. Another helpful possible work for more complete quintessence models would be to find KS teleparallel $F(T)$ solutions from a scalar field source. We may also do this type of works for phantom energy models with scalar field. We can also extend this specific study for quantized scalar fields as in ref \cite{palia2022KS}, by hoping to find similar $F(T)$ solutions to those in this paper. There are possibly arduous work, but full of hope.

\section*{Acknowledgements}

AL is supported by an Atlantic Association of Research in Mathematical Sciences (AARMS) fellowship. Thanks to A.A. Coley and R.J. van den Hoogen for their useful and constructive comments.

\section*{Abbreviations}

\noindent The following abbreviations are used in this manuscript:\\
\noindent 
\begin{tabular}{@{}ll}
AL & Alexandre Landry\\
DE & Differential Equation\\
EoS & Equation of State \\
Eqn & Equation \\
FE & Field Equation\\
GR & General Relativity \\
KV & Killing Vector \\
NGR & New General Relativity \\
TEGR & Teleparallel Equivalent of General Relativity \\
\end{tabular}

\vspace*{0.5cm}






\begin{thebibliography}{999}

\bibitem{Aldrovandi_Pereira2013} Aldrovandi, R. \& Pereira, J.G. \textit{Teleparallel Gravity, An Introduction}, Springer, 2013, \href{https://link.springer.com/book/10.1007/978-94-007-5143-9}{Link}.

\bibitem{Bahamonde:2021gfp} Bahamonde, S., Dialektopoulos, K., Escamilla-Rivera, C., Farrugia, G., Gakis, V., Hendry, M., Hohmann, M., Said, J.L., Mifsud, J. \& Di Valentino, E., Teleparallel Gravity: From Theory to Cosmology, \textit{Report Progress Physics} \textbf{2023}, \textit{86}, 026901  \href{https://arxiv.org/abs/2106.13793}{[ArXiv:2106.13793 [gr-qc]]}, \href{https://iopscience.iop.org/article/10.1088/1361-6633/ac9cef}{Link}.

\bibitem{Krssak:2018ywd} Krssak, M., van den Hoogen, R., Pereira, J., Boehmer, C. \& Coley, A., Teleparallel Theories of Gravity: Illuminating a Fully Invariant Approach, \textit{Classical And Quantum Gravity} \textbf{2019}, \textit{36}, 183001, \href{https://arxiv.org/abs/1810.12932}{[arXiv:1810.12932 [gr-qc]]}, \href{https://iopscience.iop.org/article/10.1088/1361-6382/ab2e1f}{Link}.

\bibitem{chinea1988symmetries} Chinea, F., Symmetries in tetrad theories, \textit{Classical and Quantum Gravity} \textbf{1988}, \textit{5}, 135, \href{https://iopscience.iop.org/article/10.1088/0264-9381/5/1/018}{Link}.

\bibitem{estabrook1996moving} Estabrook, F. \& Wahlquist, H., Moving frame formulations of 4-geometries having isometries, \textit{Classical and Quantum Gravity} \textbf{1996}, \textit{13}, 1333, \href{https://iopscience.iop.org/article/10.1088/0264-9381/13/6/008}{Link}.

\bibitem{papadopoulos2012locally} Papadopoulos, G. \& Grammenos, T., Locally homogeneous spaces, induced Killing vector fields and applications to Bianchi prototypes, \textit{Journal of Mathematical Physics} \textbf{2012}, \textit{53}, 072502 , \href{https://arxiv.org/abs/1106.3897}{[arXiv:1106.3897 [gr-qc]]}, \href{https://pubs.aip.org/aip/jmp/article/53/7/072502/94967/Locally-homogeneous-spaces-induced-Killing-vector}{Link}.


\bibitem{MCH} McNutt, D.D., Coley, A.A. \& van den Hoogen, R.J., A frame based approach to computing symmetries with non-trivial isotropy groups, \textit{Journal of Mathematical Physics} \textbf{2023}, \textit{64}, 032503, \href{https://arxiv.org/abs/2302.11493}{[arXiv:2302.11493 [gr-qc]]}, \href{https://pubs.aip.org/aip/jmp/article/64/3/032503/2881713/A-frame-based-approach-to-computing-symmetries}{Link}.

\bibitem{olver1995equivalence} Olver, P. \textit{Equivalence, invariants and symmetry}, Cambridge University Press, 1995, \href{https://books.google.ca/books/about/Equivalence_Invariants_and_Symmetry.html?id=YuTzf61HILAC&redir_esc=y}{Link}.


\bibitem{Ferraro:2006jd} Ferraro, R. \& Fiorini, F., Modified teleparallel gravity: Inflation without an inflation, \textit{Physical Review D} \textbf{2007}, \textit{75}, 084031, \href{https://arxiv.org/abs/gr-qc/0610067}{[arXiv:gr-qc/0610067]}, \href{https://journals.aps.org/prd/abstract/10.1103/PhysRevD.75.084031}{Link}.

\bibitem{Ferraro:2008ey} Ferraro, R. \& Fiorini, F. On Born-Infeld Gravity in Weitzenbock spacetime, \textit{Physical Review D} \textbf{2008}, \textit{78}, 124019, \href{https://arxiv.org/abs/0812.1981}{[arXiv:0812.1981 [gr-qc]]}, \href{https://journals.aps.org/prd/abstract/10.1103/PhysRevD.78.124019}{Link}.

\bibitem{Linder:2010py} Linder, E., Einstein's Other Gravity and the Acceleration of the Universe, \textit{Physical Review D} \textbf{2010}, \textit{81}, 127301, [Erratum: \textit{Physical Review D}, \textbf{2010}, \textit{82}, 109902], \href{https://arxiv.org/abs/1005.3039}{[ArXiv:1005.3039 [gr-qc]]}, \href{https://journals.aps.org/prd/abstract/10.1103/PhysRevD.81.127301}{Link}.

\bibitem{Lucas_Obukhov_Pereira2009} Lucas, T.G., Obukhov, Y. \& Pereira, J.G., Regularizing role of teleparallelism, \textit{Physical Review D} \textbf{2009}, \textit{80}, 064043, \href{https://arxiv.org/abs/0909.2418}{[arXiv:0909.2418 [gr-qc]]}, \href{https://journals.aps.org/prd/abstract/10.1103/PhysRevD.80.064043}{Link}.

\bibitem{Krssak_Pereira2015} Krssak, M. \& Pereira, J.G., Spin Connection and Renormalization of Teleparallel Action, \textit{The European Physical Journal C} \textbf{2015}, \textit{75}, 519, \href{https://arxiv.org/abs/1504.07683}{[arXiv:1504.07683 [gr-qc]]}, \href{https://link.springer.com/article/10.1140/epjc/s10052-015-3749-2}{Link}.



\bibitem{kayashi} Hayashi, K. \& Shirafuji, T., New general relativity, \textit{Physical Review D} \textbf{1979}, \textit{19}, 3524, \href{https://journals.aps.org/prd/abstract/10.1103/PhysRevD.19.3524}{Link}.

\bibitem{beltranngr} Jimenez, J.B. \& Dialektopoulos, K.F., Non-Linear Obstructions for Consistent New General Relativity, \textit{Journal of Cosmology and Astroparticle Physics} \textbf{2020} 01, 018, \href{https://arxiv.org/abs/1907.10038}{[arXiv:1907.10038 [gr-qc]]}, \href{https://iopscience.iop.org/article/10.1088/1475-7516/2020/01/018}{Link}.

\bibitem{bahamondengr} Bahamonde, S., Blixt, D., Dialektopoulos, K.F. \& Hell A., Revisiting Stability in New General Relativity, \textbf{2024}, preprint: \href{https://arxiv.org/abs/2404.02972}{[arXiv:2404.02972 [gr-qc]]}. 



\bibitem{heisenberg1} Heisenberg, L., Review on $f(Q)$ Gravity, \textbf{2023}, preprint: \href{https://arxiv.org/abs/2309.15958}{[arXiv:2309.15958 [gr-qc]]}.

\bibitem{heisenberg2} Heisenberg, L., Hohmann, M. \& Kuhn, S., Cosmological teleparallel perturbations, \textbf{2023}, preprint: \href{https://arxiv.org/abs/2311.05495}{[arXiv:2311.05495 [gr-qc]]}.

\bibitem{faithman1} Flathmann, K. \& Hohmann, M., Parametrized post-Newtonian limit of generalized scalar-nonmetricity theories of gravity, \textit{Physical Review D} \textbf{2022}, \textit{105}, 044002, \href{https://arxiv.org/abs/2111.02806}{[arXiv:2311.02806 [gr-qc]]}, \href{https://journals.aps.org/prd/abstract/10.1103/PhysRevD.105.044002}{Link}.

\bibitem{hohmannfq} Hohmann, M., General covariant symmetric teleparallel cosmology, \textit{Physical Review D} \textbf{2021}, \textit{104}, 124077, \href{https://arxiv.org/abs/2109.01525}{[arXiv:2109.01525 [gr-qc]]}, \href{https://journals.aps.org/prd/abstract/10.1103/PhysRevD.104.124077}{Link}.


\bibitem{jimeneztrinity} Jimenez, J.B., Heisenberg, L. \& Koivisto, T.S., The Geometrical Trinity of Gravity, \textit{Universe} \textbf{2019}, \textit{5}(7), 173, \href{https://arxiv.org/abs/1903.06830}{[arXiv:1903.06830 [gr-qc]]}, \href{https://www.mdpi.com/2218-1997/5/7/173}{Link}.

\bibitem{nakayama} Nakayama, Y., Geometrical trinity of unimodular gravity, \textit{Classical and Quantum Gravity} \textbf{2023}, \textit{40}, 125005, \href{https://arxiv.org/abs/2209.09462}{[arXiv:2209.09462 [gr-qc]]}, \href{https://iopscience.iop.org/article/10.1088/1361-6382/acd100/meta}{Link}.


\bibitem{ftqgravity} Xu, Y., Li, G., Harko, T. \& Liang, S.-D., $f(Q,T)$ gravity, \textit{The European Physical Journal C} \textbf{2019}, \textit{79}, 708, \href{https://arxiv.org/abs/1908.04760}{[arXiv:1908.04760 [gr-qc]]}, \href{https://link.springer.com/article/10.1140/epjc/s10052-019-7207-4}{Link}.  

\bibitem{ftqspecial} Maurya, D.C., Yesmakhanova, K., Myrzakulov, R. \& Nugmanova, G., Myrzakulov. $F(T,Q)$ gravity: cosmological implications and constraints, \textbf{2024}, preprint: \href{https://arxiv.org/abs/2404.09698}{[arXiv:2404.09698 [gr-qc]]}.

\bibitem{frqspecial} Maurya, D.C., Yesmakhanova, K., Myrzakulov, R. \& Nugmanova, G., Myrzakulov, FLRW Cosmology in Myrzakulov $F(R,Q)$ Gravity, \textbf{2024}, preprint: \href{https://arxiv.org/abs/2403.11604}{[arXiv:2403.11604 [gr-qc]]}.

\bibitem{frtspecial} Maurya, D.C. \& Myrzakulov, R., Exact Cosmology in Myrzakulov Gravity, \textbf{2024}, preprint: \href{https://arxiv.org/abs/2402.02123}{[arXiv:2402.02123 [gr-qc]]}.

\bibitem{frttheory} Harko, T., Lobo, F.S.N., Nojiri, S. \& Odintsov, S.D., $f(R,T)$ gravity, \textit{Physical Review D} \textbf{2011}, \textit{84}, 024020, \href{https://arxiv.org/abs/1104.2669}{[ArXiv:1104.2669 [gr-qc]]}, \href{https://journals.aps.org/prd/abstract/10.1103/PhysRevD.84.024020}{Link}.


\bibitem{golov1} Golovnev, A. \& Guzman, M.-J., Approaches to spherically symmetric solutions in $f(T)$-gravity, \textit{Universe} \textbf{2021}, \textit{7} (5), 121, \href{https://arxiv.org/abs/2103.16970}{[ArXiv:2103.16970 [gr-qc]]}, \href{https://www.mdpi.com/2218-1997/7/5/121}{Link}.

\bibitem{golov2} Golovnev, A., Issues of Lorentz-invariance in $f(T)$-gravity and calculations for spherically symmetric solutions, \textit{Classical and Quantum Gravity} \textbf{2021}, \textit{38}, 197001, \href{https://arxiv.org/abs/2105.08586}{[ArXiv:2105.08586 [gr-qc]]}, \href{https://iopscience.iop.org/article/10.1088/1361-6382/ac2136}{Link}.

\bibitem{golov3} Golovnev, A. \& Guzman, M.-J., Bianchi identities in $f(T)$-gravity: Paving the way to confrontation with astrophysics, \textit{Physics Letter B} \textbf{2020}, \textit{810}, 135806, \href{https://arxiv.org/abs/2006.08507}{[ArXiv:2006.08507 [gr-qc]]}, \href{https://www.sciencedirect.com/science/article/pii/S0370269320306092?via%3Dihub}{Link}.

\bibitem{debenedictis} DeBenedictis, A., Iliji\'c, S. \& Sossich, M., On spherically symmetric vacuum solutions and horizons in covariant $f(T)$ gravity theory, \textit{Physical Review D} \textbf{2022}, \textit{105}, 084020, \href{https://arxiv.org/abs/2202.08958}{[ArXiv:2202.08958 [gr-qc]]}, \href{https://journals.aps.org/prd/abstract/10.1103/PhysRevD.105.084020}{Link}.

\bibitem{SSpaper} Coley, A.A., Landry, A., van den Hoogen, R.J. \& McNutt, D.D., Spherically symmetric teleparallel geometries, \textit{The European Physical Journal C} \textbf{2024}, \textit{84}, 334, \href{https://arxiv.org/abs/2402.07238}{[ArXiv:2402.07238 [gr-qc]]}, \href{https://link.springer.com/article/10.1140/epjc/s10052-024-12629-5}{Link}.

\bibitem{TdSpaper} Coley, A.A., Landry, A., van den Hoogen, R.J. \& McNutt, D.D., Generalized Teleparallel de Sitter geometries, \textit{The European Physical Journal C} \textbf{2023}, \textit{83}, 977, \href{https://arxiv.org/abs/2307.12930}{[ArXiv:2307.12930 [gr-qc]]}, \href{https://link.springer.com/article/10.1140/epjc/s10052-023-12150-1}{Link}. 

\bibitem{baha1} Bahamonde, S. \& Camci, U., Exact Spherically Symmetric Solutions in Modified Teleparallel gravity, \textit{Symmetry} \textbf{2019}, \textit{11}, 1462, \href{https://arxiv.org/abs/1911.03965v2}{[ArXiv:1911.03965 [gr-qc]]}, \href{https://www.mdpi.com/2073-8994/11/12/1462}{Link}. 

\bibitem{awad1} Awad, A., Golovnev, A., Guzman, M.-J. \& El Hanafy, W., Revisiting diagonal tetrads: New Black Hole solutions in $f(T)$-gravity, \textit{The European Physical Journal C} \textbf{2022}, \textit{82}, 972, \href{https://arxiv.org/abs/2207.00059}{[ArXiv:2207.00059 [gr-qc]]}, \href{https://link.springer.com/article/10.1140/epjc/s10052-022-10939-0}{Link}.

\bibitem{bahagolov1} Bahamonde, S., Golovnev, A., Guzm\'an, M.-J., Said, J.L. \& Pfeifer, C., Black Holes in $f(T,B)$ Gravity: Exact and Perturbed Solutions, \textit{Journal of Cosmology and Astroparticle Physics} \textbf{2022}, \textit{01} 037, \href{https://arxiv.org/abs/2110.04087}{[ArXiv:2110.04087 [gr-qc]]}, \href{https://iopscience.iop.org/article/10.1088/1475-7516/2022/01/037}{Link}. 

\bibitem{baha6} Bahamonde, S., Faraji, S., Hackmann, E. \& Pfeifer, C., Thick accretion disk configurations in the Born-Infeld teleparallel gravity, \textit{Physical Review D} \textbf{2022}, \textit{106}, 084046, \href{https://arxiv.org/abs/2209.00020}{[ArXiv:2209.00020 [gr-qc]]}, \href{https://journals.aps.org/prd/abstract/10.1103/PhysRevD.106.084046}{Link}.

\bibitem{nashed5} Nashed, G.G.L., Quadratic and cubic spherically symmetric black holes in the modified teleparallel equivalent of general relativity: Energy and thermodynamics, \textit{Classical and Quantum Gravity} \textbf{2021}, \textit{38}, 125004, \href{https://arxiv.org/abs/2105.05688}{[ArXiv:2105.05688 [gr-qc]]}, \href{https://iopscience.iop.org/article/10.1088/1361-6382/abf89b}{Link}. 

\bibitem{pfeifer2} Pfeifer, C. \& Schuster, S., Static spherically symmetric black holes in weak $f(T)$-gravity, \textit{Universe} \textbf{2021}, \textit{7}, 153, \href{https://arxiv.org/abs/2104.00116v2}{[ArXiv:2104.00116 [gr-qc]]}, \href{https://www.mdpi.com/2218-1997/7/5/153}{Link}.

\bibitem{elhanafy1} El Hanafy, W. \& Nashed, G.G.L., Exact Teleparallel Gravity of Binary Black Holes, \textit{Astrophysical Space Science} \textbf{2016}, \textit{361}, 68, \href{https://arxiv.org/abs/1507.07377}{[ArXiv:1507.07377 [gr-qc]]}, \href{https://link.springer.com/article/10.1007/s10509-016-2662-y}{Link}.

\bibitem{benedictis3} Aftergood, J. \& DeBenedictis, A., Matter Conditions for Regular Black Holes in $f(T)$ Gravity, \textit{Physical Review D} \textbf{2014}, \textit{90}, 124006, \href{https://arxiv.org/abs/1409.4084}{[ArXiv:1409.4084 [gr-qc]]}, \href{https://journals.aps.org/prd/abstract/10.1103/PhysRevD.90.124006}{Link}.

\bibitem{baha10} Bahamonde, S., Doneva, D.D., Ducobu, L., Pfeifer, C. \& Yazadjiev, S.S., Spontaneous Scalarization of Black Holes in Gauss-Bonnet Teleparallel Gravity, \textit{Physical Review D} \textbf{2023}, \textit{107}, 10, 104013, \href{https://arxiv.org/abs/2212.07653}{[ArXiv:2212.07653 [gr-qc]]}, \href{https://journals.aps.org/prd/abstract/10.1103/PhysRevD.107.104013}{Link}.

\bibitem{baha4} Bahamonde, S., Ducobu, L. \& Pfeifer, C., Scalarized Black Holes in Teleparallel Gravity, \textit{Journal of Cosmology and Astroparticle Physics} \textbf{2022}, \textit{04}(04), 018, \href{https://arxiv.org/abs/2201.11445v2}{[ArXiv:2201.11445 [gr-qc]]}, \href{https://iopscience.iop.org/article/10.1088/1475-7516/2022/04/018}{Link}.

\bibitem{calza} Calza, M. \& Sebastiani, L., A class of static spherically symmetric solutions in $f(T)$-gravity, \textit{The European Physical Journal C} \textbf{2024}, \textit{84}, 476, \href{https://arxiv.org/abs/2309.04536}{[arXiv:2309.04536 [gr-qc]]}, \href{https://link.springer.com/article/10.1140/epjc/s10052-024-12801-x}{Link}.

\bibitem{nonvacSSpaper} A. Landry, Static spherically symmetric perfect fluid solutions in teleparallel F(T) gravity, \textit{Axioms} \textbf{2024}, \textit{13} (5), 333, \href{https://arxiv.org/abs/2405.09257}{[arXiv:2402.09257 [gr-qc]]}, \href{https://www.mdpi.com/2075-1680/13/5/333}{Link}.



\bibitem{leon1} Leon, G. \& Roque, A.A., qualitative analysis of Kantowski-Sachs metric in a generic class of $f(R)$ models,\textit{ Journal of Cosmology and Astroparticle Physics}, \textbf{2014}, \textit{05}, 032, \href{https://arxiv.org/abs/1308.5921}{[arXiv:1308.5921 [gr-qc]]}, \href{https://iopscience.iop.org/article/10.1088/1475-7516/2014/05/032}{Link}.

\bibitem{KScurvature} Shaikh, A.A. \& Chakraborty, D., Curvature properties of Kantowski-Sachs metric, \textit{Journal of Geometry and Physics}, \textit{160}, \textbf{2021}, 103970, \href{https://www.sciencedirect.com/science/article/abs/pii/S0393044020302400}{Link}.

\bibitem{KSanisotropic} Oliveira-Neto, G., Canedo, D.L. \& Monerat, G.A., An anisotropic Kantowski-Sachs universe with radiation, dust and a phantom fluid, \textit{Brazilian Journal of Physics} \textbf{2022}, \textit{52}, 130, \href{https://arxiv.org/abs/2109.12229}{[arXiv:2109.12229 [gr-qc]]}, \href{https://link.springer.com/article/10.1007/s13538-022-01137-0}{Link}.


\bibitem{Rodrigues2015} Rodrigues, M. E., Kpadonou, A.V., Rahaman, F., Oliveira, P.J. \& Houndjo, M.J.S., Bianchi type-I, type-III and Kantowski-Sachs solutions in $f(T)$ gravity, \textit{Astrophysics and Space Science} \textbf{2015}, \textit{357}, 129, \href{https://arxiv.org/abs/1408.2689}{[arXiv:1408.2689 [gr-qc]]}, \href{https://link.springer.com/article/10.1007/s10509-015-2358-8}{Link}.

\bibitem{Amir2015} Amir, M.J. \& Yussouf, M., Kantowski-Sachs Universe Models in $f(T)$ Theory of Gravity, \textit{International Journal of Theoretical Physics} \textbf{2015}, \textit{54}, 2798, \href{https://arxiv.org/abs/1502.00777}{[arXiv:1502.00777 [gr-qc]]}, \href{https://link.springer.com/article/10.1007/s10773-015-2517-2}{Link}.


\bibitem{leon2} Leon, G. \& Paliathanasis, A., Anisotropic spacetimes in $f(T,B)$ theory II: Kantowski-Sachs Universe, \textit{The European Physical Journal Plus} \textbf{2022}, \textit{137}, 855, \href{https://arxiv.org/abs/2207.08570}{[arXiv:2207.08570 [gr-qc]]}, \href{https://link.springer.com/article/10.1140/epjp/s13360-022-03083-x}{Link}.

\bibitem{leon3} Leon, G. \& Paliathanasis, A., Anisotropic spacetimes in $f(T,B)$ theory III: LRS Bianchi III Universe, \textit{The European Physical Journal Plus} \textbf{2022}, \textit{137}, 927, \href{https://arxiv.org/abs/2207.08571}{[arXiv:2207.08571 [gr-qc]]}, \href{https://link.springer.com/article/10.1140/epjp/s13360-022-03091-x}{Link}.

\bibitem{frtkssol1} Vinutha, T., Niharika, K. \& Sri Kavya, K., The Study of Kantowski-Sachs Perfect Fluid Cosmological Model in Modified Gravity, \textit{Astrophysics} \textbf{2023}, \textit{66}, 64, \href{https://arxiv.org/abs/2301.01163}{[arXiv:2301.01163 [gr-qc]]}, \href{https://link.springer.com/article/10.1007/s10511-023-09771-5}{Link}.

\bibitem{frtkssol2} Samanta, G.C., Kantowski-Sachs Universe Filled with Perfect Fluid in $f(R,T)$ Theory of Gravity, \textit{International Journal of Theoretical Physics} \textbf{2013}, \textit{52}, 2647, \href{https://link.springer.com/article/10.1007/s10773-013-1556-9}{Link}.

\bibitem{fqks1} Dimakis, N., Roumeliotis, M., Paliathanasis, A. \& Christodoulakis, T., Anisotropic Solutions in Symmetric Teleparallel $f(Q)$-theory: Kantowski-Sachs and Bianchi III LRS Cosmologies, \textit{The European Physical Journal C} \textbf{2023}, \textit{83}, 794, \href{https://arxiv.org/abs/2304.04419}{[arXiv:2304.04419 [gr-qc]]}, \href{https://link.springer.com/article/10.1140/epjc/s10052-023-11964-3}{Link}.

\bibitem{fqks2} Millano, A.D., Dialektopoulos, K., Dimakis, N., Giacomini, A., Shababi, H., Halder, A. \& Paliathanasis, A., Kantowski-Sachs and Bianchi III dynamics in $f(Q)$-gravity, \textit{Physical Review D} \textbf{2024}, \textit{109}, 124044, \href{https://arxiv.org/abs/2403.06922}{[arXiv:2403.06922 [gr-qc]]}, \href{https://journals.aps.org/prd/abstract/10.1103/PhysRevD.109.124044}{Link}.




\bibitem{palia2022KS} Paliathanasis, A., Classical and Quantum Cosmological Solutions in Teleparallel Dark Energy with Anisotropic Background Geometry, \textit{Symmetry} \textbf{2022}, \textit{14} (10), 1974, \href{https://arxiv.org/abs/2209.08817}{[arXiv:2209.08817 [gr-qc]]}, \href{https://www.mdpi.com/2073-8994/14/10/1974}{Link}.

\bibitem{palia2023KS} Paliathanasis, A., Kantowski-Sachs cosmology in scalar-torsion theory, \textit{The European Physical Journal C} \textbf{2023}, \textit{83}, 213, \href{https://arxiv.org/abs/2302.09608}{[arXiv:2302.09608 [gr-qc]]} , \href{https://link.springer.com/article/10.1140/epjc/s10052-023-11342-z}{Link}.

\bibitem{hawkingellis1} Hawking, S.W. \& Ellis, G.F.R., \textit{The Large Scale Structure of Space-Time}, Cambridge University Press, 2010, \href{https://www.cambridge.org/core/books/large-scale-structure-of-spacetime/1E6B961EC9878EDDBBD6AC0AF031CC93}{Link}.

\bibitem{coleybook} Coley, A.A., \textit{Dynamical systems and cosmology}, Kluwer Academic, Dordrecht, 2003, \href{https://link.springer.com/book/10.1007/978-94-017-0327-7}{Link}.


\bibitem{steinhardt1} Zlatev, I., Wang, L. \& Steinhardt, P., Quintessence, Cosmic Coincidence, and the Cosmological Constant, \textit{Physical Review Letters}, \textbf{1999}, \textit{82} (5), 896, \href{https://arxiv.org/abs/astro-ph/9807002}{[arXiv:astro-ph/9807002 [astro-ph]]}, \href{https://journals.aps.org/prl/abstract/10.1103/PhysRevLett.82.896}{Link}.

\bibitem{steinhardt2} Steinhardt, P., Wang, L. \& Zlatev, I., Cosmological tracking solutions, \textit{Physical Review D}, \textbf{1999}, \textit{59} (12): 123504, \href{https://arxiv.org/abs/astro-ph/9812313}{[arXiv:astro-ph/9812313 [astro-ph]]}, \href{https://journals.aps.org/prd/abstract/10.1103/PhysRevD.59.123504}{Link}

\bibitem{caldwell1} Caldwell, R.R., A phantom menace? Cosmological consequences of a dark energy component with super-negative equation of state, \textit{Physics Letters B}, \textbf{2002}, \textit{545}, 23, \href{https://arxiv.org/abs/astro-ph/9908168}{[arXiv:astro-ph/9908168 [astro-ph]]}, \href{https://www.sciencedirect.com/science/article/abs/pii/S0370269302025893?via%3Dihub}{Link}.

\bibitem{farnes} Farnes, J.S., A Unifying Theory of Dark Energy and Dark Matter: Negative Masses and Matter Creation within a Modified $\Lambda$CDM Framework, \textit{Astronomy \& Astrophysics} \textbf{2018}, \textit{620}, A92, \href{https://arxiv.org/abs/1712.07962}{[arXiv:1712.07962 [physics.gen-ph]]}, \href{https://www.aanda.org/articles/aa/full_html/2018/12/aa32898-18/aa32898-18.html}{Link}.

\bibitem{baumframpton} Baum, L. \& Frampton, P.H., Turnaround in Cyclic Cosmology, \textit{Physical Review Letters} \textbf{2007}, \textit{98}, 071301, \href{https://arxiv.org/abs/hep-th/0610213}{[arXiv:hep-th/0610213 [hep-th]]}, \href{https://journals.aps.org/prl/abstract/10.1103/PhysRevLett.98.071301}{Link}

\bibitem{attractor} Duchaniya, L.K., Gandhi, K. \&  Mishra, B., Attractor behavior of $f(T)$ modified gravity and the cosmic acceleration, \textit{Physics of the Dark Universe} \textbf{2024}, \textit{44}, 101464, \href{https://arxiv.org/abs/2303.09076}{[arXiv:2303.09076 [gr-qc]]}, \href{https://www.sciencedirect.com/science/article/pii/S2212686424000438?via%3Dihub}{Link}.

\bibitem{Kofinas} Kofinas, G., Leon, G. \& Saridakis, E.N., Dynamical behavior in $f(T,T_G)$ cosmology, \textit{Classical and Quantum Gravity} \textbf{2014}, \textit{31}, 175011, \href{https://arxiv.org/abs/1404.7100}{[arXiv:1404.7100 [gr-qc]]}, \href{https://iopscience.iop.org/article/10.1088/0264-9381/31/17/175011}{Link}.


\end{thebibliography}
\end{document}